\documentclass[fleqn,reqno]{amsart}

\usepackage{amscd,amssymb,amsmath,graphicx,verbatim}
\usepackage[pdftex]{hyperref}
\usepackage{textcomp}
\usepackage[OT1,T1]{fontenc}
\usepackage{mathabx}

\newtheorem{theorem}{Theorem}

\theoremstyle{definition}
\newtheorem{definition}{Definition}

\theoremstyle{remark}

\begin{document}

\title[Projective Representations of the Inhomogeneous Hamilton
Group]{Projective Representations of the Inhomogeneous Hamilton
Group: Noninertial Symmetry in Quantum Mechanics}
\author{S G Low}
\address{S. G. Low, www.stephen-low.net}
\email{stephen.low@alumni.utexas.edu}
\author{P D Jarvis}
\address{P. D. Jarvis, School of Mathematics and Physics, University
of Tasmania}
\email{Peter.Jarvis@utas.edu.au}
\author{R Campoamor-Stursberg}
\address{R. Campoamor-Stursberg, I.M.I-U.C.M}
\curraddr{I.M.I-U.C.M, Plaza de Ciencias 3, E-28040 Madrid, Spain}
\email{otto\_campoamor@mat.ucm.es}
\label{I1}\label{I2}\label{I3}
\date{June 7, 2011}
\begin{abstract}
Symmetries in quantum mechanics are realized by the projective representations
of the Lie group as physical states are defined only up to a phase.
A cornerstone theorem shows that these representations are equivalent
to the unitary representations of the central extension of the group.
The formulation of the inertial states of special relativistic quantum
mechanics as the projective representations of the inhomogeneous
Lorentz group, and its nonrelativistic limit in terms of the Galilei
group, are fundamental examples. Interestingly, neither of these
symmetries includes the Weyl-Heisenberg group; the hermitian representations
of its algebra are the Heisenberg commutation relations that are
a foundation of quantum mechanics. The Weyl-Heisenberg  group
is a one dimensional central extension of the abelian group and
its unitary representations are therefore a particular projective
representation of the abelian group of translations on phase space. A
theorem involving the automorphism group shows that the maximal
symmetry that leaves invariant the Heisenberg commutation relations
are essentially projective representations of the inhomogeneous
symplectic group. In the nonrelativistic domain, we must also have
invariance of Newtonian time. This reduces the symmetry group to
the inhomogeneous Hamilton group that is a local noninertial symmetry
of Hamilton's equations. The projective representations of these
groups are calculated using the Mackey theorems for the general
case of a nonabelian normal subgroup.
\end{abstract}
\maketitle

\section{Introduction}

Projective representations are required in quantum mechanics as
the physical states in quantum mechanics are rays. A ray is an equivalence
class of states in a Hilbert space that are defined up to a phase.
A cornerstone theorem states that any projective representation
of a Lie group is equivalent to the unitary representations of the
central extension of the group.\ \ The central extension of a connected\footnote{Connected
means that all elements of the group are path connected to the identity}
group is unique and simply connected.\ \ \ Levi's theorem states
that every simply connected Lie group is equivalent to the semidirect
product of a semi-simple group and a solvable normal group \cite{barut}.\ \ Mackey's
theorems provide the method of determining the unitary irreducible
representations of a general class of semidirect product groups
\cite{mackey}. We review in Section 2 these theorems that are fundamental
to the application of symmetry groups in quantum mechanics and enable
us to compute the projective representations of a very general class
of connected Lie groups. 

The inertial states of special relativistic quantum mechanics are
given by the projective representations of the inhomogeneous Lorentz
group \cite{wigner},\cite{Weinberg1}. The inhomogeneous connected
Lorentz group is the semidirect product, $\mathcal{I}\mathcal{L}(
1,n) \simeq \mathcal{L}( 1,n) \otimes _{s}\mathcal{A}( n) $ where
$\mathcal{A}( n) \simeq (\mathbb{R}^{n},+)$ is the solvable abelian
normal subgroup and the connected semisimple group is $\mathcal{L}(
1,n) $\footnote{$\mathcal{L}( 1,n) $ is the connected component
of $\mathcal{O}( 1,n) $.}. Its central extension, denoted by the
inverted caret, does not have an algebraic extension and is therefore
equal to its universal cover, denoted by a bar, $\widecheck{\mathcal{I}\mathcal{L}}(
1,n) \simeq \overline{\mathcal{I}\mathcal{L}}( 1,n) $. For $n=3$,
this is the Poincar\'e group $\overline{\mathcal{I}\mathcal{L}}(
1,3) \simeq \mathcal{S}\mathcal{L}( 2,\mathbb{C}) \otimes _{s}\mathcal{A}(
4) $ where $\overline{\mathcal{L}}( 1,3) \simeq \mathcal{S}\mathcal{L}(
2,\mathbb{C}) $\footnote{We note as a case where the symmetry group
is not connected that this may be extended to the groups $\mathcal{I}\mathcal{O}(
1,n) $ and $\mathcal{I}\mathcal{S}\mathcal{O}( 1,n) $ that have
the discrete symmetries parity-time-reversal, $(1,P,T,\mathrm{PT})$
and $(1,\mathrm{PT})$ respectively where $P$ is parity and $T$ is
time-reversal. These groups are not connected and have 2 and 4 components
respectively.\ \ While the central extension of a group that is
not connected is\ \ not necessarily unique, it turns out that, for
$\mathcal{I}\mathcal{S}\mathcal{O}( 1,n) $ it is unique and is given
in terms of the $\mathcal{S}pin( n) $ group that is the unique cover
of $\mathcal{S}\mathcal{O}( 1,n) $.\ \ On the other hand, $\mathcal{O}(
1,n) $ has 8 non-isomorphic covers that include the ${\mathcal{P}in}^{\pm
}( 1,n) $ group. [5], [6]}\ \ \cite{DeWitt},\cite{Azcarraga}

This is a truly remarkable and beautiful application of symmetry
in physics. Starting simply with the inhomogeneous Lorentz symmetry
of special relativity and the quantum mechanical condition that
physical states are rays in a Hilbert space, the general group theory
theorems result in the Hilbert spaces of the irreducible representations
of the inertial states of special relativistic quantum mechanics
labeled by the eigenvalues of the hermitian representation of the
Casimir operators. For massive states, these eigenvalues are mass
and spin where spin takes on integral and half integral values.
The half integral values associated with fermions are a consequence
of the central extension that is the Poincar\'e group. The central
extension is a direct consequence of physical states being rays
that are equivalence classes of states in a Hilbert space related
by a quantum phase.\ \ Thus the quantum phase leads directly to
the existence of fermion states. 

However, there is no mention of the Weyl-Heisenberg group; the Heisenberg
commutation relations, that are fundamental to quantum mechanics,
are the hermitian representation of the algebra for the unitary
representations of the Weyl-Heisenberg group.\ \ \ 

This is not a a consequence of special relativity as the Weyl-Heisenberg
group does not appear in the nonrelativistic formulation either.
The non-relativistic In\"on\"u-Wigner contraction of the inhomogeneous
Lorentz group is given by the inhomogeneous Euclidean group $\mathcal{I}\mathcal{E}(
n) \simeq \mathcal{E}( n) \otimes _{s}\mathcal{A}( n+1) $. $\mathcal{E}(
n) $ is the homogeneous group, $\mathcal{E}( n) \simeq \mathcal{S}\mathcal{O}(
n) \otimes _{s}\mathcal{A}( n) $, that is the non-relativistic limit
of $\mathcal{L}( 1,n) $ and is parameterized by rotations and velocity
\cite{inonu2}.\ \ The central extension of $\mathcal{I}\mathcal{E}(
n) $ admits a one parameter algebraic extension, the generator of
which is mass. This algebraic central extension defines the Galilei
group $\mathcal{G}a( n) $ whose cover is the full central extension,\ \ $\widecheck{\mathcal{I}\mathcal{E}}(
n) \simeq  \overline{\mathcal{G}a}( n) \simeq \overline{\mathcal{E}}(
n) \otimes _{s}\mathcal{A}( n+2) $ \cite{inonu2},\cite{Voisin}.
In this case, the group theory results in representations that include
inertial states with nonzero mass and energy with momentum diagonal
that are elements of a Hilbert space of square integrable functions
over $\mathbb{R}^{n}$. (For $n=3$, this is the usual 3-momentum.)\ \ Again,
there is no mention of the Weyl-Heisenberg group. 

Let us put aside these relativistic considerations for a moment
and instead consider a phase space $\mathbb{P}$ that has a symplectic
2-form $\omega $ with a symmetry group $\mathcal{S}p( 2n) $. The
elements of the group may depend on the location in phase space
and therefore the symmetry is local.\ \ Diffeomorphisms $\phi :\mathbb{P}\rightarrow
\mathbb{P}$ that leave invariant the symplectic metric, $\phi ^{*}(
\omega ) =\omega $, are the usual position-momentum canonical transformations
of classical mechanics.\ \ Locally, the Jacobian of the diffeomorphsims
must be an element of the symplectic group.\ \ If $\mathbb{P}\simeq
\mathbb{R}^{2n}$, then there is also a translational symmetry and
the symmetry group is $\mathcal{I}\mathcal{S}p( 2n) \simeq \mathcal{S}p(
2n) \otimes _{s}\mathcal{A}( 2n) $.\ \ 

Consider the projective representations of $\mathcal{I}\mathcal{S}p(
2n) $ that by the fundamental theorem, are the unitary representations
of its central extension, $\mathcal{I}\widecheck{\mathcal{S}p}( 2n)
\simeq \overline{\mathcal{S}p}( 2n) \otimes _{s}\mathcal{H}( n)
$. The Weyl-Heisenberg group $\mathcal{H}( n) $ is a one parameter
central extension of the abelian group $\mathcal{A}( 2n) $. This
is the Weyl-Heisenberg group for which the commutators of the hermitian
representation of the algebra are precisely the Heisenberg momentum-position
commutation relations.\ \ It is a direct consequence of the physical
states being rays\ \ that are equivalence classes of states in the
Hilbert phase up to a phase that this noncommutative structure arises\footnote{Note
the quote from Dirac at the beginning of the Discussion section.
}\cite{Dirac 2}.

If one instead were to consider the projective representation of
the abelian group by itself, then the central extension $\widecheck{\mathcal{A}}(
2n) $ is required. This central extension is in general $n( 2n-1)
$ dimensional. It is the presence of the symplectic homogeneous
group that constrains it to precisely the Weyl-Heisenberg group.\ \ In
fact, an automorphism theorem (Theorem 5) that we will review in
the next section, constrains a semidirect product with $\mathcal{H}(
n) $ as a normal subgroup to be a group homomorphic to a subgroup
of the automorphism group of the Weyl-Heisenberg group.\ \ The connected
component of the automorphism group is just the inhomogeneous symplectic
group with an addition conformal multiplicative term, ${\mathcal{A}ut}_{\mathcal{H}(
n) }\simeq \mathcal{D}\widecheck{\mathcal{S}p}( 2n) $ where $\mathcal{D}\mathcal{S}p(
2n) \simeq \mathcal{D}\otimes _{s}\mathcal{I}\mathcal{S}p( 2n) $
and $\mathcal{D}\simeq (\mathbb{R}^{+},\times )$. One can show that
$\mathcal{D}\widecheck{\mathcal{S}p}( 2n) \simeq \mathcal{D}\otimes
_{s}\mathcal{I}\widecheck{\mathcal{S}p}( 2n) $ and therefore it is the
symplectic group that constrains the central extension to be the
Weyl-Heisenberg group. Furthermore, this is the maximal group that
leads to the Wey-Heisenberg group as a result of a central extension.

The above considerations also apply to extended phase space with
the addition of time and energy degrees of freedom. This results
in a symplectic 2-form that may be put in the form\ \ \ 
\[
\omega =\zeta _{\alpha ,\beta }d z^{\alpha }d z^{\beta }=\delta
_{i,j}d p^{i}\wedge d q^{j}+d t \wedge d \varepsilon .
\]

\noindent As there is no relativistic line element to distinguish
time, this is just a symplectic manifold with an additional two
degrees of freedom. If $\mathbb{P}\simeq \mathbb{R}^{2n+2}$, the
symmetry is now $\mathcal{I}\mathcal{S}p( 2n+2) $.\ \ The maximal
symmetry for which the central extension results in the Weyl-Heisenberg
group is $\mathcal{D}\mathcal{S}p( 2n+2) $ with $\mathcal{D}\widecheck{\mathcal{S}p}(
2n+2) \simeq \mathcal{D}\otimes _{s}\overline{\mathcal{S}p}( 2n+2)
\otimes _{s}\mathcal{H}( n+1) $. The hermitian representations of
the algebra now, at least formally, include the time, energy commutation
relations.\ \ \ \ 

We now have the situation where we have a symmetry group whose projective
representations result in a Weyl-Heisenberg group that gives us
the Heisenberg commutation relations. But as mentioned, we do not
have a relativistic line element to define invariant time.\ \ \ For
special relativity, this is just the Minkowski line element left
invariant by the Lorentz group. In the nonrelativistic limit, it
reduces to Newtonian time 
\[
d \tau ^{2}= d t^{2}-\frac{1}{c^{2}} d q^{2} \operatorname*{\rightarrow
}\limits_{c\rightarrow \infty } d t^{2}
\]

The invariance group for $d t^{2}$ on an $m$ dimensional space is
the affine group $\mathcal{I}\mathcal{G}\mathcal{L}( m-1,\mathbb{R})
$ $\\mathrm{cite}{\mathrm{Glimore2}}$.\ \ If one also requires length
to be invariant, it reduces to the inhomogeneous Euclidean group
$\mathcal{I}\mathcal{E}( m) $ that is the nonrelativistic limit
of the inhomogeneous Lorentz group discussed above. 

In this paper, we are going to focus on the nonrelativistic case,
for which $d t^{2}$ is invariant, along with the maximal invariance
group for the Heisenberg commutation relations\footnote{Comments
on the relativistic case are given in the Discusssion section.}.
The intersection of the affine group and the homogeneous group\ \ $\mathcal{D}\otimes
_{s}\mathcal{S}p( 2n+2) $ is the group $\mathcal{H}\mathcal{S}p(
2n) \simeq \mathcal{S}p( 2n) \otimes _{s}\mathcal{H}( n) $.\ \ This
group is acting on the tangent space and this Weyl-Heisenberg group
is parameterized by force, velocity and power. To understand more
clearly the meaning of this symmetry, consider diffeomorphisms $\phi
:\mathbb{P}\rightarrow \mathbb{P}$ that leave both $\omega $ and
$d t^{2}$ invariant: $\phi ^{*}( \omega ) =\omega $ and $\phi ^{*}(
d t^{2}) =d t^{2}$. Then the Jacobian of this diffeomorphism must
be an element of\ \ $\mathcal{H}\mathcal{S}p( 2n) $.\ \ The semidirect
group properties\ \ imply that the diffeomorphism can be written
as\ \ $\phi =\varphi \circ \widetilde{\varphi }$ where both $\varphi
$ and $\widetilde{\varphi }$ are diffeomorphisms. The Jacobian of $\widetilde{\varphi
}$ take their values in $\mathcal{S}p( 2n) $ and are therefore just
the usual canonical transformations. The Jacobian of $\varphi $
take their values in $\mathcal{H}( n) $ (that is parameterized by
force, velocity and power) and this can be shown to be equivalent
to Hamilton's equations \cite{Low7}. In fact, $\mathcal{H}\mathcal{S}p(
2n) $ is the maximal local symmetry of Hamilton's equations.\ \ Requiring
also invariance of length reduces the $\mathcal{H}\mathcal{S}p(
2n) $ to the Hamilton group that is defined to be $\mathcal{H}a(
n) \simeq \mathcal{S}\mathcal{O}( n) \otimes _{s}\mathcal{H}( n)
$.

The projective representations of $\mathcal{I}\mathcal{H}\mathcal{S}p(
2n) \simeq \mathcal{H}\mathcal{S}p( 2n) \otimes \mathcal{A}( 2n+2)
$ are the unitary representations of its central extension
\[
\mathcal{I}\mathcal{H}\widecheck{\mathcal{S}p}( 2n) \simeq \overline{\mathcal{H}\mathcal{S}p}(
2n) \otimes _{s}\mathcal{H}( n+1) \simeq \overline{\mathcal{S}p}(
2n) \otimes _{s}\mathcal{H}( n) \otimes _{s}\mathcal{H}( n+1) 
\]

The hermitian representation of the algebra of $\mathcal{H}( n+1)
$ are the Heisenberg commutation relations.\ \ The group $\mathcal{H}(
n) $, on the other hand, is parameterized by force, velocity and
power and acts, along with $\mathcal{S}p( 2n) $, on the tangent
or cotangent space of $\mathbb{P}$.

The projective representations of $\mathcal{I}\mathcal{H}a( n) \simeq
\mathcal{H}a( n) \otimes \mathcal{A}(2n+2)$ are also the unitary
representations of its central extension that we will show is
\[
\mathcal{I}\widecheck{\mathcal{H}a}( n) \simeq \overline{\mathcal{H}a}(
n) \otimes _{s}\mathcal{H}( n+1) \simeq \overline{\mathcal{S}O}(
n) \otimes _{s}\mathcal{H}( n) \otimes _{s}\left( \mathcal{H}( n+1)
\otimes \mathcal{A}( 2) \right) 
\]

The above comments on $\mathcal{H}( n) $ and $\mathcal{H}( n+1)
$ apply here also. The group $\overline{\mathcal{G}a}( n) $ is an
inertial subgroup of $\mathcal{I}\widecheck{\mathcal{H}a}( n) $. This
group has both the symmetry of Galilean relativity as its inertial
special case and the Weyl-Heisenberg group that leads to the Heisenberg
commutation relations.

The outline of the paper is as follows. Section two presents the
mathematical framework that provides the theorems that enable the
projective representations of a very general class of connected
Lie groups to be determined in a fully tractable manner. These theorems
are fundamental to symmetry in quantum mechanics as physical states
are rays in a Hilbert space and therefore require projective representations\ \ \ 

Section 3 studies the central extension of the inhomogeneous Hamilton
group. The Hamilton group is of interest as it is a symmetry of
the classical nonrelativistic Hamilton's equations from which it
derives its name.\ \ This local symmetry is valid for both inertial
and noninertial states. The Euclidean subgroup parameterized by
rotations and velocity is the inertial subgroup.\ \ The projective
representations of the inhomogeneous Euclidean subgroup result in
the inertial states of nonrelativistic quantum mechanics. This motivates
the study of the projective representations of the inhomogeneous
Hamilton group for the more general noninertial states.\ \ 

 Section 4 of this paper studies the projective representations
of the inhomogeneous Hamilton group, $\mathcal{I}\mathcal{H}a( n)
$ as a global symmetry. This requires us to use the full power of
the theorems of Section 2, including the nonabelian normal subgroup
case of the Mackey theorems. As preliminary steps we use the theorems
to compute the representations of the Weyl-Heisenberg group and
the Hamilton group as these are required in the full result.\ \ As
the theorems are required in their general form, these calculations
are illustrative of how the projective representations of a general
class of groups can be computed. The physical application of these
representations are then discussed. 

\section{Mathematical framework}\label{PH: Section: Mathematical
Framework}

In this section we review a set of theorems that enable us to compute
the projective representations for a general class of connected
Lie groups. We refer to the cited literature for full proofs as
several of the theorems are deep. Our purpose here is to assemble
the theorems in a form that, when applied together, provide us with
tractable tools to compute this general class of representations
that includes the much studied unitary\footnote{$\mathcal{S}\mathcal{U}(
n) \simeq \widecheck{\mathcal{S}\mathcal{U}}( n) $ and is semisimple
so the theorems are trivial in this case. $\mathcal{I}\mathcal{L}(
1,n) $ has an abelian normal subgroup for which\ \ the theorems
simplify.\ \ As these are the most studied groups in physics, these
more general theorems generally have not received the attention
they deserve as the key theorems of symmetry in quantum mechanics}
and inhomogeneous Lorentz group cases.\ \ These are fundamental
theorems on the application of symmetry groups in quantum mechanics.

Our notation for a semidirect product is $\mathcal{G}\simeq \mathcal{K}\otimes
_{s}\mathcal{N}$ where $\mathcal{N}$ is the normal subgroup and
$\mathcal{K}$ is the homogeneous subgroup such that $\mathcal{K}\cap
\mathcal{N}=\text{\boldmath $e$}$ and $\mathcal{G}\simeq \mathcal{N}
\mathcal{K}$\footnote{The notation for a semidirect product varies
with many authors placing the normal subgroup on the left and the
reader should be sure to check conventions. Note that the semidirect
product differs from a direct product in that multiplication of
the normal group from the right is $\mathcal{G}\simeq (\varsigma
_{\mathcal{N}}\mathcal{K})\mathcal{N}$ where $\varsigma _{h}:\mathcal{G}\rightarrow
\mathcal{G}:g \mapsto h g h^{-1}$, $h\in \mathcal{N}$.}. A semidirect
product is right associative in the sense that $\mathcal{D}\simeq
(\mathcal{A}\otimes _{s}\mathcal{B})\otimes _{s}\mathcal{C}$ implies
that $\mathcal{D}\simeq \mathcal{A}\otimes _{s}(\mathcal{B}\otimes
_{s}\mathcal{C})$ and so brackets can be removed. However $\mathcal{D}\simeq
\mathcal{A}\otimes _{s}(\mathcal{B}\otimes _{s}\mathcal{C})$ does
not necessarily imply $\mathcal{D}\simeq (\mathcal{A}\otimes _{s}\mathcal{B})\otimes
_{s}\mathcal{C}$ as $\mathcal{B}$ is not necessarily a normal subgroup
of $\mathcal{A}$.\ \ 
\begin{definition}

An algebraic central extension of a Lie algebra $g$ is the Lie algebra
$\widecheck{g}$ that satisfies the following short exact sequence where
$z$ is the maximal abelian algebra that is central in $\widecheck{g}$,\label{PH:
def: alg central extension}
\begin{equation}
\text{\boldmath $0$}\rightarrow z\rightarrow \widecheck{g}\rightarrow
g\rightarrow \text{\boldmath $0$} .
\end{equation}
\end{definition}

\noindent where $\text{\boldmath $0$}$ is the trivial algebra. Suppose
$\{X_{a}\}$ is a basis of the Lie algebra $g$ with commutation relations
$[X_{a},X_{b}]=c_{a,b}^{c}X_{c}$, $a,b=1,...r$.\ \ Then an algebraic
central extension is a maximal set of central abelian generators
$\{A_{\alpha }\}$, where $\alpha ,\beta ,... =1,..m$,\ \ such that
\begin{equation}
\left[ A_{\alpha },A_{\beta }\right] =0,\ \ \ \ \left[ X_{a},A_{\alpha
}\right] =0,\ \ \ \ \left[ X_{a},X_{b}\right] =c_{a,b}^{c}X_{c}+c_{a,b}^{\alpha
}A_{\alpha }
\end{equation}

\noindent The basis $\{X_{a},A_{\alpha }\}$ of the centrally extended
Lie algebra must also satisfy the Jacobi identities. The Jacobi
identities\ \ constrain the admissible central extensions of the
algebra. The choice\ \ $X_{a}\mapsto X_{a}+A_{a}$ will always satisfy
these relations and this trivial case is exclude.\ \ The algebra
$\widecheck{g}$ constructed in this manner is equivalent to the central
extension of $g$ given in Definition 1.
\begin{definition}

The central extension of a connected Lie group $\mathrm{\mathcal{G}}$
is the\ \ Lie group $\widecheck{\mathcal{G}}$ that satisfies the following
short exact sequence where $\mathrm{\mathcal{Z}}$ is a maximal abelian
group that is central in $\widecheck{\mathcal{G}}$
\begin{equation}
\text{\boldmath $e$}\rightarrow \mathcal{Z}\rightarrow \widecheck{\mathcal{G}}\overset{\pi
}{\rightarrow }\mathcal{G}\rightarrow \text{\boldmath $e$} .
\end{equation}
\end{definition}

The abelian group $\mathcal{Z}$ may always be written as the direct
product $\mathcal{Z}\simeq \mathcal{A}( m) \otimes \mathbb{A}$ of
a connected continuous abelian Lie group $\mathcal{A}( m) \simeq
(\mathbb{R}^{m},+)$ and a discrete abelian group $\mathbb{A}$ that
may have a finite or countable dimension \cite{bargmann},\cite{mackey2}.

The exact sequence may be decomposed into an exact sequence for
the {\itshape topological} central extension and the {\itshape algebraic}
central extension,
\begin{equation}
\text{\boldmath $e$}\rightarrow \mathbb{A}\rightarrow \overline{\mathcal{G}}\overset{\pi
\mbox{}^{\circ}}{\rightarrow }\mathcal{G}\rightarrow \text{\boldmath
$e$} \text{\boldmath $,$}\text{\boldmath $\ \ $}\text{\boldmath
$e$}\rightarrow \mathcal{A}( m) \rightarrow \widecheck{\mathcal{G}}\overset{\widetilde{\pi
}}{\rightarrow }\overline{\mathcal{G}}\rightarrow \text{\boldmath
$e$}.
\end{equation}

\noindent where $\pi =\pi \mbox{}^{\circ}\circ \widetilde{\pi }$. The
first exact sequence defines the universal cover where $\mathbb{A}\simeq
\ker  \pi \mbox{}^{\circ}$ is the fundamental homotopy group. All
of the groups is in the second sequence are simply connected and
therefore may be defined by the exponential map of the central extension
of the algebra given by Definition 1. In other words, the full central
extension may be computed by determining the universal covering
group of the algebraic central extension.

A ray $\Psi $ is the equivalence class of states $|\psi \rangle
$ in a Hilbert space $\text{\boldmath $\mathrm{H}$}$ up to a phase,
\begin{equation}
\Psi =\left\{ e^{i \omega }\left| \psi \right\rangle   |\omega \in
\mathbb{R}\right\} ,\ \ \ \left| \psi \right\rangle  \in \text{\boldmath
$\mathrm{H}$}
\end{equation}

\noindent Note that the physical quantities that are the square
of the modulus depend only on the ray
\[
|\left( \Psi _{\beta },\Psi _{\alpha }\right) |^{2}={\left| \left\langle
\psi _{\beta }|\psi _{\alpha }\right\rangle  \right| }^{2}
\]

\noindent for all $|\psi _{\gamma }\rangle \in \Psi $.\ \ 
\begin{definition}

A projective representation $\varrho $ of a symmetry group $\mathcal{G}$$\text{}$
is the maximal representation such that\ \ for\ \ $|{\widetilde{\psi
}}_{\gamma }\rangle =\varrho ( g) |\psi _{\gamma }\rangle $, the
modulus is invariant\ \ ${|\langle {\widetilde{\psi }}_{\beta }|{\widetilde{\psi
}}_{\alpha }\rangle |}^{2}={|\langle \psi _{\beta }|\psi _{\alpha
}\rangle |}^{2}$ for all\ \ $|\psi _{\gamma }\rangle ,|{\widetilde{\psi
}}_{\gamma }\rangle \in \Psi $. \label{PH: def: proj representation}
\end{definition}
\begin{theorem}

{\bfseries (Wigner, Weinberg)} Any projective representation of
a Lie symmetry group $\mathcal{G}$ on a separable Hilbert space
is equivalent to a representation that is either linear and unitary
or anti-linear and anti-unitary. Furthermore, if $\mathcal{G}$ is
connected, the projective representations are equivalent to a representation
that is linear and unitary \cite{Wignerb},\cite{Weinberg1}.\label{PH:
theorem: Wigner unitary projective}
\end{theorem}

This is the generalization of the well known theorem that the ordinary
representation of any compact group is equivalent to a representation
that is unitary.\ \ For a projective representation, the phase degrees
of freedom of the central extension enables the equivalent linear
unitary or antilinear antiunitary representation to be constructed
for this much more general class of Lie groups that admit representations
on separable Hilbert spaces.\ \ (A proof of the theorem is given
in Appendix A of Chapter 2 of \cite{Weinberg1}).\ \ The groups that
this theorem applies to include the non-compact inhomogeneous\ \ Euclidean,
Lorentz, Hamilton or unitary groups that are studied in this paper.
\begin{theorem}

{\bfseries (Bargmann, Mackey)} The projective representations of
a connected Lie group $ \mathcal{G}$ are equivalent to the ordinary
unitary representations of its central extension $\widecheck{\mathcal{G}}$
\cite{bargmann},\cite{mackey2}.\label{PH: theorem: proj rep is unitary
CE}
\end{theorem}

Theorem 1 states that are all projective representations are equivalent
to a projective representation that is unitary. A phase is the unitary
representation of a central abelian subgroup. Therefore, the maximal
representation is given in terms of the central extension of the
group. Appendix A shows how this definition is equivalent to the
formulation of a projective representation as an ordinary unitary
representation that is defined {\itshape up to a phase},\ \ $\varrho
( \gamma ^{\prime }) \varrho ( \gamma ) =e^{i \theta }\varrho (
\gamma ^{\prime }\gamma ) $\cite{bargmann},\cite{mackey2}.\ \ \ \ \ 
\begin{theorem}

Let $\mathcal{G}$,$\mathcal{H}$ be Lie groups and $\pi :\mathcal{G}\rightarrow
\mathcal{H}$ be a homomorphism. Then, for every unitary representation
$\widetilde{\varrho }$ of $\mathcal{H}$ there exists a degenerate unitary
representation $\varrho $ of $\mathcal{G}$ defined by $\varrho =\widetilde{\varrho
}\circ \pi $. Conversely, for every degenerate unitary representation
of a Lie group $\mathcal{G}$ there exists a Lie subgroup $\mathcal{H}$
and a homomporphism $\pi :\mathcal{G}\rightarrow \mathcal{H}$ where
$\ker ( \pi ) \neq \text{\boldmath $e$}$ such that $\varrho =\widetilde{\varrho
}\circ \pi $\ \ where $\widetilde{\varrho }$ is a unitary representation
of $\mathcal{H}$.\label{PH: theorem: degenerate reps}
\end{theorem}

Noting that a representation is a homomorphism, This theorem follows
straightforwardly from the properties of homomorphisms. As a consequence,
the set of degenerate representations of a group is characterized
by its set of normal subgroups. A {\itshape faithful} representation
is the case that the representation is an isomorphism.
\begin{theorem}

{\bfseries (Levi) }Any simply connected\ \ Lie group is equivalent
to the semidirect product of a semisimple group and and a maximal
solvable normal subgroup \cite{barut}.\label{PH: theorem: Levi}
\end{theorem}

As the central extension of any connected group is simply connected,
the problem of computing the projective representations of a group
always can be reduced to computing the unitary irreducible representations
of a semidirect product group with a semisimple homogeneous group
and a solvable normal subgroup.\ \ The unitary irreducible representations
of the semisimple groups are known and the solvable groups that
we are interested in turn out to be the semidirect product of abelian
groups.
\begin{theorem}

Any semidirect product group $\mathcal{G}\simeq \mathcal{K}\otimes
_{s}\mathcal{N}$ is a subgroup of a group homomorphic to the group
of automorphisms of the normal subgroup, $\mathcal{G}\subset {\mathcal{A}ut}_{\mathcal{N}}$
\cite{barut}.\label{PH: theorem: automorphisms semid-direct}
\end{theorem}

This theorem places constraints on the admissible semidirect product
groups that have a given normal subgroup.\ \ For example, the automorphism
group of the abelian group is
\begin{equation}
 {\mathcal{A}ut}_{\mathcal{A}( m) }\simeq \mathbb{Z}_{2}\otimes
_{s}\mathcal{D}\otimes _{s}\overline{\mathcal{G}\mathcal{L}}( m,\mathbb{R})
\otimes _{s}\mathcal{A}( m)  , \mathcal{D}\simeq \left( \mathbb{R}^{+},\times
\right) ,%
\label{PH: Abelain automorphism group}
\end{equation}

\noindent whereas the automorphism group of the Weyl-Heisenberg
group $\mathcal{H}( m) $ is \cite{folland},\cite{Low8}
\begin{equation}
 {\mathcal{A}ut}_{\mathcal{H}( m) }\simeq \mathbb{Z}_{2}\otimes
_{s}\mathcal{D}\otimes _{s}\overline{\mathcal{S}p}( 2m) \otimes
_{s}\mathcal{H}( m) .%
\label{PH: Weyl-Heisenberg automorphism group}
\end{equation}

This means that there does not exist a semidirect product of the
form $\mathcal{S}\mathcal{O}( 2m) \otimes _{s}\mathcal{H}( m) $
as $\mathcal{S}\mathcal{O}( 2m) $ is not a subgroup of $\mathcal{S}p(
2m) $. On the other hand, the semidirect product $\mathcal{H}a(
n) =\mathcal{S}\mathcal{O}( m) \otimes _{s}\mathcal{H}( m) $ is
admissible as $\mathcal{S}\mathcal{O}( m) \subset \mathcal{S}p(
2m) $.$\text{}$

\subsection{Mackey theorems for the representations of semidirect
product groups}

The Mackey theorems are valid for a general class of topological
groups but we will only require the more restricted case $\mathcal{G}\simeq
\mathcal{K}\otimes _{s}\mathcal{N}$ where the group $\mathcal{G}$
and subgroups $\mathcal{K},\mathcal{N}$ are smooth Lie groups.\ \ The
central extension of any connected Lie group is simply connected
and therefore generally has the form of a semidirect product due
to Theorem 3 (Levi)\footnote{The semisimple or solvable group may
be trivial in which case the semidirect product is trivial}.\ \ Theorem
5 further constrains the possible homogeneous groups $\mathcal{K}$
of the semidirect product given the normal subgroup $\mathcal{N}$.

 The first Mackey theorem is the induced representation theorem
that gives a method of constructing a unitary representation of
a group (that is not necessarily a semidirect product group) from
a unitary representation of a closed subgroup. The second theorem
gives a construction of certain representations of a certain subgroup
of a semidirect product group from which the complete set of unitary
irreducible representations of the group can be induced. This theorem
is valid for the general case where the normal subgroup $\mathcal{N}$
is a nonabelian group. In the special case where the normal subgroup
$\mathcal{N}$ is abelian, the theorem may be stated in a simpler
form.
\begin{theorem}

{\bfseries {\upshape (Mackey)}} {\bfseries {\upshape Induced representation
theorem.}} Suppose that $\mathcal{G}$ is a Lie group and $\mathcal{H}$
is a Lie subgroup, $\mathcal{H}\subset \mathcal{G}$ such that $\mathbb{K}\simeq
\mathcal{G}/\mathcal{H}$ is a homogeneous space with a natural projection\ \ $\pi
:\mathcal{G}\rightarrow \mathbb{K}$, an invariant measure and a
canonical section {\upshape $\Theta :\mathbb{K}\rightarrow \mathcal{G}:\mathrm{k}\mapsto
g$} such that\ \ $\pi \circ \Theta \mathrm{=}{\mathrm{Id}}_{\mathbb{K}}$
where ${\mathrm{Id}}_{\mathbb{K}}$ is the identity map on $\mathrm{\mathbb{K}}$.
Let $\rho $ be a unitary representation of $\mathcal{H}$ on the
Hilbert space ${\text{\boldmath $\mathrm{H}$}}^{\rho }$:\label{PH:
theorem: Mackey induction theorem}
\[
\rho ( h) :{\text{\boldmath $\mathrm{H}$}}^{\rho }\rightarrow {\text{\boldmath
$\mathrm{H}$}}^{\rho }:\left| \varphi \right\rangle  \mapsto \left|
\widetilde{\varphi }\right\rangle  =\rho ( h) \left| \varphi \right\rangle
,\ \ h\in \mathcal{H}.
\]

\noindent Then a unitary representation $\mathrm{\varrho }$ of a
Lie group $\mathrm{\mathcal{G}}$ on the Hilbert space ${\text{\boldmath
$\mathrm{H}$}}^{\mathrm{\varrho }}$,
\[
\varrho ( g) :{\text{\boldmath $\mathrm{H}$}}^{\varrho }\rightarrow
{\text{\boldmath $\mathrm{H}$}}^{\varrho }:\left| \psi \right\rangle
\mapsto \left| \widetilde{\psi }\right\rangle  =\varrho ( g) \left|
\psi \right\rangle  ,\ \ g\in \mathcal{G},
\]

\noindent may be induced from the representation $\mathrm{\rho }$
of $\mathrm{\mathcal{H}}$\ \ by defining
\begin{equation}
\widetilde{\psi }( \mathrm{k}) =\left( \varrho ( g) \psi \right) \left(
\mathrm{k}\right) =\rho ( \mathit{g\mbox{}^{\circ}}) \psi ( g^{-1}\mathrm{k})
,\text{}\ \ \ \mathit{g\mbox{}^{\circ}}={\Theta ( \mathrm{k}) }^{-1}
g \Theta ( g^{-1}\mathrm{k}) 
\end{equation}

\noindent where the Hilbert space on which the induced representation
$\mathrm{\varrho }$ acts is given by ${\text{\boldmath $\mathrm{H}$}}^{\varrho
}\simeq {\text{\boldmath $\mathrm{L}$}}^{2}( \mathbb{K},H^{\rho
}) $ \cite{mackey}, \cite{barut}.
\end{theorem}

The proof is straightforward given that the section $\Theta $ exists
by showing first that $g \mbox{}^{\circ}\in \ker ( \pi ) \simeq
\mathcal{H}$ and therefore $\rho ( g \mbox{}^{\circ}) $ is well
defined.
\begin{definition}

{\bfseries (}{\bfseries {\itshape Little groups}}{\bfseries )} Let
$\mathcal{G}=\mathcal{K}\otimes _{s}\mathcal{N}$ be a semidirect
product. Let $[\xi ]\in {\text{\boldmath $U$}}_{\mathcal{N}}$ where\ \ ${\text{\boldmath
$U$}}_{\mathcal{N}}$ denotes the unitary dual whose elements are
equivalence classes of unitary representations of $\mathcal{N}$
on a Hilbert space ${\text{\boldmath $\mathrm{H}$}}^{\xi }$.\ \ Let
$\rho $ be a unitary representation of a subgroup $\mathcal{G}\mbox{}^{\circ}=\mathcal{K}\mbox{}^{\circ}\otimes
_{s}\mathcal{N}$ on the Hilbert space ${\text{\boldmath $\mathrm{H}$}}^{\xi
}$ such that $\rho _{|\mathcal{N}}=\xi $. The {\itshape little}
{\itshape groups} are the set of maximal subgroups $\mathcal{K}^{\circ
}$ such that $\rho $ exists on the corresponding {\itshape stabilizer}
$\mathcal{G}^{\circ }\simeq \mathcal{K}\mbox{}^{\circ}\otimes _{s}\mathcal{N}$
and satisfies the fixed point equation\label{PH: defn: little group}
\begin{equation}
 {\widehat{\varsigma }}_{\rho ( k) }[ \xi ] =\left[ \xi \right]  , k\in
\mathcal{K}\mbox{}^{\circ}.%
\label{PH: Little group general equation}
\end{equation}
\end{definition}

\noindent In this definition the dual automorphism is defined by
\begin{equation}
\begin{array}{ll}
 \left( {\widehat{\varsigma }}_{\rho ( g) }\xi \right) \left( h\right)
& =\rho ( g)  \rho ( h) {\rho ( g) }^{-1}=\rho ( g h g^{-1}) =\xi
(  \varsigma _{g}h) 
\end{array}%
\label{PH: automorphisms of little group}
\end{equation}

\noindent for all $g\in \mathcal{G}\mbox{}^{\circ}$ and $h\in \mathcal{N}$.\ \ The
equivalence classes of the unitary representations of $\mathcal{N}$
are defined by
\begin{equation}
 \left[ \xi \right] =\left\{ {\widehat{\varsigma }}_{\xi ( h) }\xi |h\in
\mathcal{N}\right\} .%
\label{PH: Abelian xi equivalence classess}
\end{equation}

\noindent A group $\mathcal{G}$ may have multiple little groups
${\mathcal{K}\mbox{}^{\circ}}_{\alpha }$ whose intersection is the
identity element only.\ \ We will generally leave the label $\alpha
$ implicit.
\begin{theorem}

{\bfseries {\upshape (Mackey)}} {\bfseries {\upshape Unitary irreducible
representations of semidirect products. }}Suppose that we have a
semidirect product Lie group {\upshape $\mathcal{G}\simeq \mathcal{K}\otimes
_{s}\mathcal{N}$}, where $\mathcal{K},\mathcal{N}$ are Lie subgroups.
Let $\mathrm{\xi }$ be the unitary irreducible representation of
$\mathrm{\mathcal{N}}$ on the Hilbert space ${\text{\boldmath $\mathrm{H}$}}^{\xi
}$.\ \ Let $\mathcal{G}\mbox{}^{\circ}\simeq \mathcal{K}\mbox{}^{\circ}\otimes
_{s}\mathcal{N}$ be a maximal stabilizer on which there exists a
representation $\rho $ on ${\text{\boldmath $\mathrm{H}$}}^{\xi
}$ such that $\rho |_{\mathcal{N}}=\xi $.\ \ \ Let $\mathrm{\sigma
}$ be a unitary irreducible representation of $\mathrm{\mathcal{K}}\mbox{}^{\circ}$
on the Hilbert space ${\text{\boldmath $\mathrm{H}$}}^{\sigma }$.
Define the representation $ \varrho \mbox{}^{\circ}=\sigma \otimes
\rho $ that acts on the Hilbert space\ \ ${\text{\boldmath $\mathrm{H}$}}^{\varrho
\mbox{}^{\circ}}\simeq {\text{\boldmath $\mathrm{H}$}}^{\sigma }\otimes
{\text{\boldmath $\mathrm{H}$}}^{\xi }$. \label{PH: theorem: Mackey
semidirect product theorem}Determine the complete set of stabilizers
and representations $\rho $\ \ and little groups that satisfy these
properties, that we label by $\alpha $,$\{{(\mathcal{G}\mbox{}^{\circ},\varrho
\mbox{}^{\circ},{\text{\boldmath $\mathrm{H}$}}^{\varrho \mbox{}^{\circ}})}_{\alpha
}\}$.\ \ If for some member of this set $\mathrm{\mathcal{G}}\text{{\upshape
$ \mbox{}^{\circ}$ $ \simeq $ $ \mathcal{G}$}}$ then for this case
the representations\ \ are\ \ {\upshape $(\mathcal{G},\varrho ,{\text{\boldmath
$\mathrm{H}$}}^{\varrho })\mathrm{\simeq }(\mathcal{G}\mbox{}^{\circ},\varrho
\mbox{}^{\circ},{\text{\boldmath $\mathrm{H}$}}^{\varrho \mbox{}^{\circ}})$}.\ \ For
the cases where the stabilizer\ \ {\upshape $\mathcal{G}\mbox{}^{\circ}$}
is a proper subgroup of $\mathrm{\mathcal{G}}$ then the unitary
irreducible representations $(\mathcal{G},\varrho ,{\text{\boldmath
$\mathrm{H}$}}^{\varrho })$ are the representations induced (using
Theorem 6) by the representations {\upshape $(\mathcal{G}\mbox{}^{\circ},\varrho
\mbox{}^{\circ},{\text{\boldmath $\mathrm{H}$}}^{\varrho \mbox{}^{\circ}})$}
of the stabilizer subgroup. The complete set of unitary irreducible
representations is the union of the representations $\cup _{\alpha
}\{{(\mathcal{G},\varrho ,{\text{\boldmath $\mathrm{H}$}}^{\varrho
})}_{\alpha }\mathrm{\}}$ over the set of all the stabilizers and
corresponding little groups \cite{mackey}.
\end{theorem}

This major result and its proof are due to Mackey \cite{mackey}.
Our focus in this paper is on applying this theorem.

\subsubsection{Abelian normal subgroup}

The theorem simplifies for special cases where the normal subgroup
$\mathcal{N}$ is an abelian group, $\mathcal{N}\simeq \mathcal{A}(
n) $.\ \ \ From Theorem 5, a semidirect product with $\mathcal{A}(
n) $ as a normal subgroup is a subgroup of a homomorphism of the
automorphism group (6).

An abelian group has the property that its unitary irreducible representations
$\xi $ are the characters acting on the Hilbert space ${\text{\boldmath
$\mathrm{H}$}}^{\xi }\simeq \mathbb{C}$,
\begin{equation}
\xi ( a)  \left| \phi \right\rangle  = e^{i a\cdot \nu }\left| \phi
\right\rangle  ,\ \ a, \nu \in \mathbb{R}^{n}.
\end{equation}

\noindent The unitary\ \ irreducible representations are labeled
by the $\nu _{i}$ that are the eigenvalues of the hermitian representation
of the basis $\{A_{i}\}$ of the abelian Lie algebra,
\begin{equation}
{\widehat{A}}_{i}\left| \phi \right\rangle  =\xi ^{\prime }( A_{i})
\left| \phi \right\rangle  = \nu _{i}\left| \phi \right\rangle 
.
\end{equation}

\noindent \ \ The equivalence classes $[\xi ]\in {\text{\boldmath
$U$}}_{\mathcal{A}( n) }$ each have a single element $[\xi ]\simeq
\xi $ as, for the abelian group, the expression (11) is trivial.
The representations $\rho $ act on ${\text{\boldmath $\mathrm{H}$}}^{\xi
}\simeq \mathbb{C}$ and are one dimensional and therefore must\ \ commute
with the $\xi $. Therefore, in equation (10),\ \ $\rho ( g) \xi
( h) {\rho ( g) }^{-1}=\xi ( h) $ and (9) simplifies to
\begin{equation}
\xi ( a) =\xi (  \varsigma _{k}a) =\xi ( k a k^{-1}) , a\in \mathcal{A}(
m) ,\ \ k\in \mathcal{K}\mbox{}^{\circ}\text{}.\ \ %
\label{PH: Little group abelian equation}
\end{equation}
\begin{theorem}

{\bfseries {\upshape (Mackey)}} {\bfseries {\upshape Unitary irreducible
representations of a semidirect product with an abelian normal subgroup.}}\ \ Suppose
that we have a semidirect product group {\upshape $\mathcal{G}\simeq
\mathcal{K}\otimes _{s}\mathcal{A}$} where $\mathcal{A}$ is abelian.
Let $\mathrm{\xi }$ be the unitary irreducible representation (that
are the characters) of $\mathrm{\mathcal{A}}$ on ${\text{\boldmath
$\mathrm{H}$}}^{\xi }\simeq \mathbb{C}$. Let $\mathcal{K}\mbox{}^{\circ}\subseteq
\mathcal{K}$ be a Little group defined by (14) with the corresponding
stabilizers\ \ $\mathcal{G}\mbox{}^{\circ}\simeq \mathcal{K}\mbox{}^{\circ}\otimes
_{s}\mathcal{A}$.\ \ Let $\mathrm{\sigma }$ be the unitary irreducible
representations of $\mathrm{\mathcal{K}}\mbox{}^{\circ}$ on the
Hilbert space ${\text{\boldmath $\mathrm{H}$}}^{\sigma }$. Define
the representation $ \varrho \mbox{}^{\circ}=\sigma \otimes \xi
$ of the stabilizer that acts on the Hilbert space\ \ ${\text{\boldmath
$\mathrm{H}$}}^{\varrho \mbox{}^{\circ}}\simeq {\text{\boldmath
$\mathrm{H}$}}^{\sigma }\otimes \mathbb{C}$.\ \ \ The theorem then
proceeds as in the case of the general Theorem 7.\label{PH: theorem:
Mackey abelian case}
\end{theorem}

\section{The central extension of the inhomogeneous Hamilton group}\label{PH:
section: CE of IHa}

Consider a $2n+2$ dimensional extended phase space manifold $\mathbb{P}\simeq
\mathbb{R}^{2n+2}$ with coordinates $(q^{i},p^{i},t,\varepsilon
)$, $i,j,...=1,...,n$,\ \ that\ \ respectively are interpreted as
position, momentum, time and energy degrees of freedom. Assume that
these are global canonical coordinates in which a symplectic two
form is $\omega =-d \varepsilon \wedge d t+\delta _{i,j}d p^{i}\wedge
d q^{i}$ and the invariant Newtonian time element is $d t^{2}$.\ \ $\mathcal{H}\mathcal{S}p(
2n) =\mathcal{S}p( 2n) \otimes _{s}\mathcal{H}( n+1) $ is the connected
symmetry group that\ \ leaves both $\omega $ and $d t^{2}$ invariant.
That is, a diffeomorphism $\varphi :\mathbb{P}\rightarrow \mathbb{P}$
for which the pullbacks satisfy $\varphi ^{*}( \omega ) =\omega
$ and $\varphi ^{*}( d t^{2}) =d t^{2}$ have Jacobians that are
elements of the local symmetry group $\mathcal{H}\mathcal{S}p( 2n)
$.\ \ This can be shown to be equivalent to the diffeomorphisms
satisfying Hamilton's equations \cite{Low7}.\ \ The subgroup of
$\mathcal{H}\mathcal{S}p( 2n) $ that leaves invariant both $d q^{2}$
and $d p^{2}$ is the Hamilton group,
\begin{equation}
\mathcal{H}a( n) \simeq  \mathcal{S}\mathcal{O}( n) \otimes _{s}\mathcal{H}(
n) .%
\label{PH: Hamilton group basic semidirect product}
\end{equation}

The Weyl-Heisenberg group is the semidirect product of two abelian
groups,
\begin{equation}
\mathcal{H}( n) \simeq \mathcal{A}( n) \otimes _{s}\mathcal{A}(
n+1) ,%
\label{PH: WH group semidirect product}
\end{equation}

\noindent that is a simply connected solvable group.\ \ In this
case, it is parameterized by velocity $v$, force $f$ and power $r$.
It is a one parameter central extension of $\mathcal{A}( 2n) $\footnote{This
is not the most general central extension of $\mathcal{A}( 2n) $.
The most general extension is $n( 2n-1) $ dimensional. }. As the
extended phase space is $\mathbb{P}\simeq \mathbb{R}^{2n+2}$, it
is also invariant under the abelian translation group $\mathcal{A}(
2n+2) $.\ \ The full symmetry group is the inhomogeneous Hamilton
group
\begin{equation}
\mathcal{I}\mathcal{H}a( n) \simeq \mathcal{H}a( n) \otimes _{s}\mathcal{A}(
2n+2) .
\end{equation}

The projective representations of the inhomogeneous Hamilton group
are given by the unitary representations of its central extension.
The central extension has been determined in \cite{Low8} to be $\mathcal{I}\widecheck{\mathcal{H}a}(
n) \simeq \overline{\mathcal{Q}\mathcal{H}a}( n) $ where $\mathcal{Q}\mathcal{H}a(
n) $ is the {\itshape quantum mechanical }Hamilton group that is
defined to be the 3 parameter algebraic central extension of $\mathcal{I}\mathcal{H}a(
n) $,
\begin{equation}
\mathcal{Q}\mathcal{H}a( n) =\ \ \mathcal{H}a( n) \otimes _{s}\left(
\mathcal{H}( n+1) \otimes \mathcal{A}( 2) \right) .%
\label{PH: QHa usual semidirect product form}
\end{equation}

\noindent In this semidirect product form, the homogeneous group
$\mathcal{K}$ is the Hamilton group and the normal subgroup is the
solvable group $\mathcal{N}\simeq \mathcal{H}( n+1) \otimes \mathcal{A}(
2) $. We will see shortly that the hermitian representation of the
algebra corresponding to the unitary representations of the Weyl-Heisenberg
subgroup $\mathcal{H}( n+1) $ are precisely the Heisenberg commutation
relations for momentum and position and energy and time.\ \ Expanding
the Hamilton group with (15) and using the right associativity of
the semidirect product, the full central extension may be written
as
\begin{equation}
\mathcal{I}\widecheck{\mathcal{H}a}( n) \simeq \overline{\mathcal{Q}\mathcal{H}a}(
n) \simeq \overline{\mathcal{S}\mathcal{O}}( n) \otimes _{s}\mathcal{H}(
n) \otimes _{s}\left( \mathcal{H}( n+1) \otimes \mathcal{A}( 2)
\right) .%
\label{PH: semidirect expansion QHA bar}
\end{equation}

\noindent This satisfies Levi's Theorem 4 where the simply connected
semisimple group is $\mathcal{K}\simeq \overline{\mathcal{S}\mathcal{O}}(
n) $ and the simply connected maximal solvable normal subgroup is
$\mathcal{N}\simeq \mathcal{H}( n) \otimes _{s}(\mathcal{H}( n+1)
\otimes \mathcal{A}( 2) )$. 

$\mathcal{Q}\mathcal{H}a( n) $ is a matrix group with elements realized
by the matrix given in Appendix B (120). Using that parameterization,
the group product for elements $\Gamma $ may be written as
\begin{equation}
\begin{array}{l}
 \Gamma ( {\mathrm{R}}^{{\prime\prime}},v^{{\prime\prime}},f^{{\prime\prime}},r^{{\prime\prime}},q^{{\prime\prime}},t^{{\prime\prime}},p^{{\prime\prime}},\varepsilon
^{{\prime\prime}},\iota ^{{\prime\prime}},s^{{\prime\prime}},u^{{\prime\prime}})
\\
 \ \ \ \ \ \ = \Gamma ( {\mathrm{R}}^{\prime },v^{\prime },f^{\prime
},r^{\prime },q^{\prime },t^{\prime },p^{\prime },\varepsilon ^{\prime
},\iota ^{\prime },s^{\prime },u^{\prime })  \Gamma ( \mathrm{R},v,f,r,q,t,p,\varepsilon
,\iota ,s,u) ,
\end{array}%
\label{PH: QHa group product}
\end{equation}

\noindent where $v,f,p,q\in \mathbb{R}^{n}$ and $r,t,\varepsilon
,\iota ,s,u\in \mathbb{R}$, and $\mathrm{R}\in \mathcal{S}\mathcal{O}(
n) $ is an $n\times n$ real matrix and\footnote{The dot denotes
the inner product $a\cdot b=a^{\mathrm{t}}b=\delta _{i,j}a^{i}b^{j}$.
The $\mathrm{t}$ denotes transpose for the matrix notation. Matrix
multiplication is implicit.}
\begin{gather*}
\begin{array}{lll}
 {\mathrm{R}}^{{\prime\prime}}={\mathrm{R}}^{\prime } \mathrm{R}\mathrm{,}
& v^{{\prime\prime}}=v^{\prime }+{\mathrm{R}}^{\prime } v, & r^{{\prime\prime}}=r^{\prime
}+r+v^{\prime }\cdot {\mathrm{R}}^{\prime }f-f^{\prime }\cdot {\mathrm{R}}^{\prime
}v, \\
 t^{{\prime\prime}}=t^{\prime }+t, & f^{{\prime\prime}}=f^{\prime
}+{\mathrm{R}}^{\prime } f, & \varepsilon ^{{\prime\prime}}=\varepsilon
^{\prime }+\varepsilon +v^{\prime }\cdot {\mathrm{R}}^{\prime }p-f^{\prime
}\cdot {\mathrm{R}}^{\prime }q+r^{\prime }t,
\end{array}
\\\begin{array}{ll}
 q^{{\prime\prime}}=q^{\prime }+{\mathrm{R}}^{\prime } q+ v^{\prime
}t, & s^{{\prime\prime}}=s+s^{\prime }+v^{\prime }\cdot {\mathrm{R}}^{\prime
}q+\frac{1}{2} t\ \ {v^{\prime }}^{2}, \\
 p^{{\prime\prime}}=p^{\prime }+{\mathrm{R}}^{\prime } p+f^{\prime
}t, & u^{{\prime\prime}}=u+u^{\prime }+f^{\prime }\cdot {\mathrm{R}}^{\prime
}p+\frac{1}{2}t\ \ {f^{\prime }}^{2},
\end{array}
\\\begin{array}{ll}
 \iota ^{{\prime\prime}}= &  \iota + \iota ^{\prime }+\frac{1}{2}\left(
\left( e^{\prime }+q^{\prime }\cdot f^{\prime }-p^{\prime }\cdot
v^{\prime }-r^{\prime }t^{\prime }\right) t- e t^{\prime }\right.
\\
   & \left. -\left( p^{\prime }-t^{\prime }f^{\prime }\right) \cdot
{\mathrm{R}}^{\prime } q+\left( q^{\prime }-t^{\prime }v^{\prime
}\right) \cdot {\mathrm{R}}^{\prime } p\right) .
\end{array}
\end{gather*}

\noindent The inverse elements of the group are
\begin{equation}
\Gamma ^{-1}( \mathrm{R},v,f,r,q,t,p,\varepsilon ,\iota ,s,u) =\Gamma
( {\mathrm{R}}^{\prime },v^{\prime },f^{\prime },r^{\prime },q^{\prime
},t^{\prime },p^{\prime },\varepsilon ^{\prime },\iota ^{\prime
},s^{\prime },u^{\prime })  %
\label{PH: QHa group inverse}
\end{equation}

\noindent where
\[
\begin{array}{lll}
 {\mathrm{R}}^{\prime }={\mathrm{R}}^{-1}\mathrm{,} & t^{\prime
}=-t,  & \varepsilon ^{\prime }=-\varepsilon +v\cdot p-f\cdot q+r
t, \\
 v^{\prime }=-{\mathrm{R}}^{-1}v, & q^{\prime }=-{\mathrm{R}}^{-1}q+
t {\mathrm{R}}^{-1}v, & s^{\prime }=-s+v\cdot q- \frac{1}{2}t v^{2},
\\
 f^{\prime }=-{\mathrm{R}}^{-1}f, & p^{\prime }=-{\mathrm{R}}^{-1}p+
t {\mathrm{R}}^{-1}f, & u^{\prime }=-u+ f\cdot p-\frac{1}{2}t f^{2},
\\
 \iota ^{\prime }=-\iota , & r^{\prime }=-r. &  
\end{array}
\]

\noindent Consider the semidirect product form given in (18). The
elements of the subgroups are
\begin{equation}
\begin{array}{l}
 \mathrm{N}( q,t,p,\varepsilon ,\iota ,s,u) =\Gamma ( 1,0,0,0,q,t,p,\varepsilon
,\iota ,s,u) \in \mathcal{N}\simeq \mathcal{H}( n+1) \otimes \mathcal{A}(
2) , \\
 \mathrm{K}( \mathrm{R},v,f,r) =\Gamma ( \mathrm{R},v,f,r,0,0,0,0,0,0,0)
\in \mathcal{K}\simeq \mathcal{H}a( n) ,
\end{array}%
\label{PH: QHa N K definitions}
\end{equation}

\noindent and
\begin{equation}
\begin{array}{l}
 \Upsilon ( q,t,p,\varepsilon ,\iota ) \simeq \mathrm{N}( q,t,p,\varepsilon
,\iota ,0,0) \in \mathcal{H}( n+1) , \\
 \mathrm{R}\simeq \mathrm{K}( \mathrm{R},0,0,0) \in \mathcal{S}\mathcal{O}(
n) , \\
 \mathrm{A}( s,u) \simeq \mathrm{N}( 0,0,0,0,0,s,u) \in \mathcal{A}(
2) , \\
 \widetilde{\Upsilon }( v,f,r) \simeq \mathrm{K}( 1,v,f,r) \in \mathcal{H}(
n) .
\end{array}%
\label{PH: subgroup elments of CE iha}
\end{equation}

\noindent The group products and inverses for these subgroups follow
directly from (20) and (21).\ \ The properties of the semidirect
product then implies that
\begin{equation}
\begin{array}{rl}
 \Gamma ( \mathrm{R},v,q,t,p,\varepsilon ,\iota ,s,u)  & =\mathrm{N}(
q,t,p,\varepsilon ,\iota ,s,u) \mathrm{K}( \mathrm{R},v)  \\
  & =\Upsilon ( q,t,p,\varepsilon ,\iota ) \mathrm{A}( s,u) \widetilde{\Upsilon
}( v,f,r) R .
\end{array}%
\label{PH: semidirect product properties qha}
\end{equation}

\noindent This can be verified explicitly using the group product
(20). 

\subsection{Lie algebra of $\mathcal{Q}\mathcal{H}a( n) $}

A general element of the algebra of $\mathcal{Q}\mathcal{H}a( n)
$ may be written as
\begin{equation}
Z=\alpha ^{i,j}J_{i,j}+v^{i}G_{i}+f^{i}F_{i}+r R+\frac{1}{\hbar
}\left(  q^{i }P_{i}+t E+p^{i}Q_{i}+\varepsilon  T\right)  +s M+u
A+\iota  I.%
\label{PH: general element of algebra}
\end{equation}

\noindent The $\alpha ^{i,j}$ are the $n( n-1) /2$ dimensionless
rotation angles, $v^{i}$ has dimensions of velocity, $f^{i}$ force,
$r$ power, $q^{i}$ position, $t$ time, $p^{i}$ momentum and $\varepsilon
$ has dimensions of energy.\ \ Next, consider the parameters of
the central extension.\ \ The parameter $s$ has dimensions of ${\mathrm{mass}}^{-1}$,
$u $ has dimensions of tension and $\iota $ is dimensionless. ($\hbar
$ is Planck's constant with the dimensions of action.) The generators
have dimensions so that a general element $Z$ is dimensionless.\ \ That
is, $J_{i,j}$ is dimensionless, $G_{i},F_{i},R$ have dimensions
of velocity${}^{-1}$, force${}^{-1}$ and power${}^{-1}$ respectively
and $P_{i},E,Q_{i},T$\ \ have dimensions of momentum, energy, position
and time respectively. The central generators $M,A$ have of dimensions
of mass and reciprocal tension and $I$ is dimensionless. 

The algebra of the Hamilton group is \cite{Low8}: 
\begin{equation}
\begin{array}{l}
 \left[ J_{i,j},J_{k,l}\right] = \delta _{i,l}J_{j,k}+\delta _{j,k}J_{i,l}
-\delta _{j,l}J_{i,k} -\delta _{i,k}J_{j,l},\ \ \left[ G_{i},F_{k}\right]
= \delta _{i,k}R, \\
 \left[ J_{i,j},G_{k}\right] =\delta _{j,k}G_{i}-\delta _{i,k}G_{j}
,\ \ \ \ \ \ \ \ \left[ J_{i,j},F_{k}\right] =\delta _{j,k}F_{i}
-\delta _{i,k}F_{j}.
\end{array}%
\label{PH: hamilton algebra}
\end{equation}

\noindent The inhomogeneous Hamilton group $\mathcal{I}\mathcal{H}a(
n) $ requires the additional nonzero commutation relations: 
\begin{equation}
\begin{array}{lll}
 \left[ J_{i,j},Q_{k}\right] =\delta _{j,k}Q_{i} -\delta _{i,k}Q_{j}
, & \left[ G_{i},Q_{k}\right] = \delta _{i,k}T, & \left[ G_{i},E\right]
=P_{i}, \\
 \left[ J_{i,j},P_{k}\right] =\delta _{j,k}P_{i} -\delta _{i,k}P_{j}
, & \left[ F_{i},P_{k}\right] =- \delta _{i,k}T, & \left[ F_{i},E\right]
=Q_{i}, \\
 \left[ R,E\right] =2T. &   &  
\end{array}%
\label{PH: inhomogeneous hamilton algebra}
\end{equation}

\noindent The above relations are the algebra for $\mathcal{I}\mathcal{H}a(
n) $. $T$ is a central generator as it commutes with all the generators.
Classically, all observers related by the inhomogeneous Hamilton
group have an invariant definition of time.\ \ The central extension
$\mathcal{Q}\mathcal{H}a( n) $ requires the additional nonzero commutation
relations
\begin{equation}
\left[ P_{i},Q_{k}\right] =\hbar  \delta _{i,k}I,\ \ \left[ E,T\right]
=-\hbar  I,\ \ \ \left[ G_{i},P_{k}\right] = \delta _{i,k}M,\ \ \left[
F_{i},Q_{k}\right] = \delta _{i,k}A.%
\label{PH: AIM central extension of IHa}
\end{equation}

\noindent $I, M, A$ are central generators as they commute with
all the other generators. $T$ is no longer central due to the nonzero
$[E,T]$ commutation relation.\ \ 

\subsection{Subgroups of $\overline{\mathcal{Q}\mathcal{H}a}( n)
$}

\subsubsection{Weyl-Heisenberg}

The quantum Hamilton group has several Weyl-Heisenberg subgroups.
One subgroup is $\mathcal{H}( n+1) $ that has an algebra spanned
by $\{P_{i},Q_{i},E,T,I\}$. The hermitian representation of the
algebra corresponding to the unitary representations of this group,
as we will see in Section 4, are the usual Heisenberg commutation
relations for position-time and energy-momentum.\ \ 

Another Weyl-Heisenberg $\mathcal{H}( n) $ subgroup is the subgroup
that is generated by $\{F_{i},G_{i},R\}$.\ \ Furthermore, there
are\ \ two Weyl-Heisenberg subgroup with an algebra generated by
$\{G_{i},P_{i},M\}$ and $\{F_{i},Q_{i},A\}$ respectively. Finally,
there are the two additional Weyl-Heisenberg subalgebras generated
by $\{G_{i},Q_{k},T\}$ and $\{F_{i},P_{i},T\}$.\ \ (Note that $\{I,M,A\}$
are central generators of the algebra of the full $\overline{\mathcal{Q}\mathcal{H}a}(
n) $ group whereas $T$ and $R$ are not.) 

\subsubsection{Hamilton}

The cover of the Hamilton group $\overline{\mathcal{H}a}( n) $ is
a subgroup of $\overline{\mathcal{Q}\mathcal{H}a}( n) $ with an
algebra with the general element
\begin{equation}
Z=\alpha ^{i,j}J_{i,j}+v^{i}G_{i}+f^{i}F_{i}+r R.%
\label{PH: Ha subalgebra of QHA}
\end{equation}

\noindent It may be written as a semidirect product in either of
the forms
\begin{equation}
\overline{\mathcal{H}a}( n) \simeq \overline{\mathcal{S}\mathcal{O}}(
n) \otimes _{s}\mathcal{H}( n) \simeq \overline{\mathcal{E}}( n)
\otimes _{s}\mathcal{A}( n+1) ,\ \ \ \overline{\mathcal{E}}( n)
\simeq \overline{\mathcal{S}\mathcal{O}}( n) \otimes _{s}\mathcal{A}(
n) .%
\label{PH: Hamilton group semidirect product form}
\end{equation}

The cover of the special orthogonal group $\overline{\mathcal{S}\mathcal{O}}(
n) $ has generators $\{J_{i,j}\}$.\ \ In the first semidirect product
form, the Weyl-Heisenberg subgroup has generators $\{G_{i},F_{i},R\}$
noted in the previous section. Alternatively, the Weyl-Heisenberg
group may be expanded as the semidirect product of abelian groups
given in (16). There are two choices for its normal subgroup $\mathcal{A}(
n+1) $. The first case, the $\mathcal{A}( n+1) $ subgroup has an
algebra spanned by\ \ $\{F_{i},R\}$ and the second case it is spanned
by $\{G_{i},R\}$.\ \ The corresponding generators for the homogenous
group $\overline{\mathcal{E}}( n) $ are $\{J_{i,j},G_{i}\}$ for
the first case and $\{J_{i,j},F_{i}\}$ for the second.

\subsubsection{Galilei}

The cover of the Galilei group may be written in several different
semidirect product forms 
\begin{equation}
\begin{array}{lll}
 \widecheck{\mathcal{I}\mathcal{E}}( n) \simeq \overline{\mathcal{G}a}(
n)  & \simeq  & \overline{\mathcal{E}}( n) \otimes _{s}\mathcal{A}(
n+2)  \\
   & \simeq  & \left( \mathcal{A}( 1) \otimes \overline{\mathcal{S}\mathcal{O}}(
n) \right) \otimes _{s}\mathcal{H}( n) \simeq \mathcal{A}( 1) \otimes
_{s}\overline{\mathcal{H}a}( n)  \\
   & \simeq  & \overline{\mathcal{S}\mathcal{O}}( n) \otimes _{s}\mathcal{A}(
1) \otimes _{s}\mathcal{H}( n) .
\end{array}%
\label{PH: Galilei group semidirect product forms}
\end{equation}

\noindent The last form is the Levi decomposition where the simply
connected semisimple homogeneous group is $\mathcal{K}\simeq $$\overline{\mathcal{S}\mathcal{O}}(
n) $ and\ \ the simply connected solvable normal subgroup is $\mathcal{N}\simeq
\mathcal{A}( 1) \otimes _{s}\mathcal{H}( n) $.

There are two distinct $\overline{\mathcal{G}a}( n) $ groups that
are subgroups of $\overline{\mathcal{Q}\mathcal{H}a}( n) $ with
general elements given by 

 $\overline{\mathcal{Q}\mathcal{H}a}( n) $ has two distinct Galilei
group subgroups\ \ $\overline{\mathcal{G}a}( n) $. General elements
of their respective Lie algebras take the form
\begin{gather}
Z_{1}=\alpha ^{i,j}J_{i,j}+v^{i}G_{i}+\frac{1}{\hbar }\left(  q^{i
}P_{i}+t E\right)  +s M.%
\label{PH: Galilean general element of algebra}
\\Z_{2}=\alpha ^{i,j}J_{i,j}+f^{i}F_{i}+\frac{1}{\hbar }\left( 
p^{i}Q_{i}+t E\right)  +u A.%
\label{PH Galilean conjugate general element of algebra}
\end{gather}

Elements of the form $Z_{1}$ generate the the usual {\itshape physical}
Galilei group. For this subgroup, written as the semidirect product
$\overline{\mathcal{E}}( n) \otimes _{s}\mathcal{A}( n+2) $, the
subgroup $\overline{\mathcal{E}}( n) $ has generators $\{J_{i,j},G_{i}\}$
and the $\mathcal{A}( n+2) $ has generators $\{P_{i},E,M\}$.\ \ \ \ Written
as the semidirect product $(\mathcal{A}( 1) \otimes \overline{\mathcal{S}\mathcal{O}}(
n) )\otimes _{s}\mathcal{H}( n) $, the subgroup $\mathcal{A}( 1)
\otimes \overline{\mathcal{S}\mathcal{O}}( n) $ has generators $\{E,J_{i,j}\}$
and the $\mathcal{H}( n) $ has generators $\{G_{i},P_{i},M\}$.\ \ 

Alternatively, elements of the form $Z_{2}$ general a different
Galilei subgroup $\overline{\mathcal{G}a}( n) $. Again, this group
may be written as the semidirect product $\overline{\mathcal{E}}(
n) \otimes _{s}\mathcal{A}( n+2) $, where the $\overline{\mathcal{E}}(
n) $ has generators $\{J_{i,j},F_{i}\}$ and the $\mathcal{A}( n+2)
$ has generators $\{Q_{i},T,A\}$. Written as the semidirect product
$(\mathcal{A}( 1) \otimes \overline{\mathcal{S}\mathcal{O}}( n)
)\otimes _{s}\mathcal{H}( n) $, the $\mathcal{A}( 1) \otimes \overline{\mathcal{S}\mathcal{O}}(
n) $ has generators $\{E,J_{i,j}\}$ and the $\mathcal{H}( n) $ has
generators $\{F_{i},Q_{i},A\}$. 

\subsection{Casimir invariants }

Any element in the center $\text{\boldmath $\mathrm{z}$}( g) $ of
the enveloping algebra $\text{\boldmath $\mathrm{e}$}( g) $ is invariant
of the algebra $g$. The Casimir invariants form a basis in the sense
that any element of the center $\text{\boldmath $\mathrm{z}$}( g)
$ may be written as a polynomial of the basis elements. The number
of $N_{c}$ Casimir invariants of $\mathcal{Q}\mathcal{H}a( n) $
and of its Galilei and Hamilton subgroups is given in the following
table:\footnote{The cover of these groups, $\overline{\mathcal{H}a}(
n) $, $\overline{\mathcal{G}a}( n) $ and $\overline{\mathcal{Q}\mathcal{H}a}(
n) $ have the same algebra and therefore the same Casimir invariants.}
\begin{table}[h]
\begin{tabular}[l]{c|ccccc}
 & $1$ & $2$ & $3$ & $4$ & $n$ \\\hline
$\mathcal{H}(n)$  & $1$ & $1$ & $1$ & $1$ & $1$ \\
$\mathcal{H}a( n)$  & $1$ & $2$ & $2$ & $3$ & $\left\lfloor
\frac{n}{2}\right\rfloor
+1$ \\
$\mathcal{G}a( n)$  & $2$ & $2$ & $3$ & $3$ & $\left\lfloor
\frac{n}{2}\right\rfloor
+2$ \\
$\mathcal{Q}\mathcal{H}a( n)$  & $4$ & $4$ & $5$ & $5$ &
$\left\lfloor \frac{n}{2}\right\rfloor +4$
\end{tabular}
\end{table}

The Casimir invariants\ \ of\ \ $\mathcal{Q}\mathcal{H}a( 3) $ have
been determined to be \cite{Low8},\cite{rutwig}
\begin{equation}
C_{1}=I, C_{2}=M, C_{3}=A, C_{4}=T^{2}-I R, C_{5}=B_{i,j}B_{i,j},%
\label{PH: Hamilton casimirs}
\end{equation}

\noindent The $B_{i,j}$ are defined, using the auxiliary invariant
$C$ as 
\begin{equation}
B_{i,j}=C J_{i,j}+D_{i,j},\ \ \ C=-A M+T^{2} -I R. %
\label{PH: noninertial spin}
\end{equation}

\noindent The $D_{i,j}$ are given by
\begin{equation}
D_{i,j}=A D_{i,j}^{1}+M D_{i,j}^{2}+R D_{i,j}^{3}+ I D_{i,j}^{4}+T
\left( D_{i,j}^{5}+D_{i,j}^{6}\right) ,%
\label{PH: noninertial spin components}
\end{equation}

\noindent where
\begin{equation}
\begin{array}{lll}
 D_{i,j}^{1}=G_{j} P_{i}-G_{i} P_{j}, & D_{i,j}^{3}=P_{i} Q_{j}-P_{j}
Q_{i} & D_{i,j}^{5}=F_{i} P_{j}-F_{j} P_{i}, \\
 D_{i,j}^{2}=F_{j} Q_{i}-F_{i} Q_{j}, & D_{i,j}^{4}=F_{i} G_{j}-F_{j}
G_{i}, & D_{i,j}^{6}=G_{i} Q_{j}-G_{j} Q_{i}.
\end{array}
\end{equation}

\noindent The Casimir invariant of $\mathcal{H}( n) $ is the central
element $I$.\ \ The Casimir invariants of\ \ $\mathcal{H}a( 3) $
subgroup are 
\begin{equation}
C_{1}=R,\ \ C_{2}= B_{i,j}B_{i,j},\ \ B_{i,j}= R J_{i,j}+F_{j}G_{i}-F_{i}G_{j}%
\label{PH: Casimir Ha}
\end{equation}

\noindent The Casimir invariants of the two $\mathcal{G}a( 3) $
subgroups are 
\begin{equation}
\begin{array}{ll}
 C_{1}=M, & C_{2}= 2M E-P^{2}, \\
 C_{3}= B_{i,j}B_{i,j}, & B_{i,j}=M J_{i,j}-G_{j}P_{i}+G_{j}P_{i},
\end{array}%
\label{PH: Casimir Ga}
\end{equation}

\noindent and
\begin{equation}
\begin{array}{ll}
 C_{1}=A, & C_{2}= 2A E-Q^{2}, \\
 C_{3}= B_{i,j}B_{i,j}, & B_{i,j}=A J_{i,j}-F_{j}Q_{i}+F_{i}Q_{j}.
\end{array}
\end{equation}

\subsection{Homomorphisms}

The groups homomorphic to $\overline{\mathcal{Q}\mathcal{H}a}( n)
$\ \ also appear in the representation theory as degenerate representations
due to Theorem 3. The homomorphisms are characterized by the normal
subgroups of the group that are the kernels of the homomorphisms.\ \ Appendix
C summarizes the homomorphic groups for the Weyl-Heisenberg, Hamilton,
Galilei and the quantum mechanical Hamilton groups.

\section{Projective representations of the inhomogeneous Hamilton
group}\label{PH: section projective reps}

The Mackey theorems may be used to compute the unitary irreducible
representations of the central extension of the inhomogeneous Hamilton
group that are required by quantum mechanics. We will need the unitary
irreducible representations of the Weyl-Heisenberg and Hamilton
group in this calculation and we therefore review these first.

A unitary representation\ \ $\varrho : \mathcal{G}\rightarrow \text{\boldmath
$U$}( \text{\boldmath $\mathrm{H}$}) : g\mapsto \varrho ( g) $\ \ satisfies
the unitary condition ${\varrho ( g) }^{-1}={\varrho ( g) }^{\dagger
}$. For a simply connected group, the group elements are $g=e^{X}$.\ \ The
representation is $\varrho ( g) =e^{ \varrho ^{\prime }( X) }$ and
the unitary condition requires $\varrho ^{\prime }( X) =-{\varrho
^{\prime }( X) }^{\dagger }$ and therefore the representation of
the algebra is anti-hermitian.\ \ The standard physics convention
is to use hermitian operators by defining $X\mapsto -i X$ so that
$\varrho ^{\prime }( X) ={\varrho ^{\prime }( X) }^{\dagger }$.\ \ This
requires an $i$ to also be inserted in the algebra commutation relations.
That is, if $[ X_{i},X_{j}]= c_{i,j}^{k}X_{k}$ then the hermitian
representation ${\widehat{X}}_{i}=\varrho ^{\prime }( X_{i}) $ satisfies
the commutation relations 
\begin{equation}
\left[  {\widehat{X}}_{i},{\widehat{X}}_{j}\right] =i c_{i,j}^{k}{\widehat{X}}_{k}.%
\label{PH: i in the algebra}
\end{equation}

\subsection{ Unitary irreducible representations of the Weyl-Heisenberg
group }\label{PH: section: WH unitary}

The unitary representations of the Weyl-Heisenberg group
\begin{equation}
 \mathcal{H}( n) \simeq \mathcal{A}( n) \otimes _{s}\mathcal{A}(
n+1) ,%
\label{PH: WH semidirect product}
\end{equation}

\noindent may be determined using the abelian Mackey Theorem 8 \cite{Major},\cite{Low}.\ \ The
Weyl-Heisenberg group product and inverse for elements $\Upsilon
( a,b,\iota ) $ is the special case (23) of (20), 
\begin{equation}
\begin{array}{l}
 \Upsilon ( a^{\prime },b^{\prime },\iota ^{\prime })  \Upsilon
( a,b,\iota ) =\Upsilon ( a^{\prime }+a,b^{\prime },+b,\iota ^{\prime
}+\iota +\frac{1}{2}\left( a^{\prime }\cdot b-b^{\prime }\cdot a\right)
) , \\
 \Upsilon ^{-1}( a,b,\iota ) =\Upsilon ( -a,-b,-\iota ) .
\end{array}%
\label{PH: WH group product}
\end{equation}

The inner automorphisms are
\begin{equation}
\varsigma _{\Upsilon ( a^{\prime },b^{\prime },\iota ^{\prime })
}\Upsilon ( a,b,\iota ) =\Upsilon ( a,b,\iota +a^{\prime }\cdot
b-b^{\prime }\cdot a) .
\end{equation}

Elements of the normal subgroup $\mathcal{A}( n+1) $ may be taken
to be either $\Upsilon ( a,0,\iota ) $ or $\Upsilon ( 0,b,\iota
) $. The corresponding elements of the homogeneous group $\mathcal{A}(
n) $ are $\Upsilon ( 0,b,0) $ or $\Upsilon ( a,0,0) $. A general
element of the Weyl-Heisenberg group may be written as 
\begin{equation}
\Upsilon ( a,b,\iota ) =\Upsilon ( a,0,\iota -\frac{1}{2}a\cdot
b) \Upsilon ( 0,b,0) =\Upsilon ( 0,b,\iota +\frac{1}{2}a\cdot b)
\Upsilon ( a,0,0) .%
\label{PH: Weyl-Heisenberg element factoring}
\end{equation}

\noindent Noting that
\begin{equation}
\zeta _{\pm }:\mathcal{H}( n) \rightarrow \mathcal{H}( n) :\Upsilon
( a,b,\iota ) \mapsto \Upsilon ( a,b,\iota \pm \frac{1}{2}a\cdot
b) ,%
\label{PH: WH isomorphism}
\end{equation}

\noindent is a group isomorphism, it may be straightforwardly shown
that either choice of the normal subgroup elements\ \ in the expressions
in (45) satisfy the properties to be the semidirect product (42).

\subsubsection{Mackey abelian semidirect product theorem}

The Mackey Theorem 8 may now be applied. We choose the normal subgroup
with elements $\Upsilon ( a,0,\iota ) \in \mathcal{A}( n+1) $. The
unitary irreducible representations $\xi $ of the abelian normal
subgroup are the phases acting on the Hilbert space ${\text{\boldmath
$\mathrm{H}$}}^{\xi }=\mathbb{C}$
\begin{equation}
\left. \xi ( \Upsilon ( a,0,\iota ) ) |\phi \right\rangle  =e^{i(
a^{i}{\widehat{A}}_{i}+\iota  \widehat{I} ) }\overset{ }{\left. |\phi \right\rangle
=e^{i( a\cdot \alpha +\iota  \lambda  ) }\overset{ }{\left. |\phi
\right\rangle  }},\ \ \overset{ }{\left. |\phi \right\rangle  }\in
\mathbb{C}.
\end{equation}

\noindent The hermitian representation of the algebra has the eigenvalues
that are given by
\begin{equation}
 {\widehat{A}}_{i}\overset{ }{\left. |\phi \right\rangle  }=\xi ^{\prime
}( A_{i}) \left| \phi \right\rangle  =\alpha _{i}\overset{ }{\left.
|\phi \right\rangle  },\ \ \ \widehat{I}\overset{ }{\left. |\phi \right\rangle
}=\xi ^{\prime }( I) \left| \phi \right\rangle  =\lambda \overset{
}{\left. |\phi \right\rangle  },\ \ 
\end{equation}

\noindent where $\alpha \in \mathbb{R}^{n}$ and $\lambda \in \mathbb{R}$.\ \ The
characters\ \ $\xi _{\alpha ,\lambda }$ are parameterized by the
eigenvalues $\alpha ,\lambda $ and the equivalence classes that
are elements of the unitary dual, $[\xi _{\alpha ,\lambda }]\in
{\text{\boldmath $U$}}_{\mathcal{A}( n+1) }\simeq \mathbb{R}^{n+1}$.\ \ Each
equivalence class has the single element $[\xi _{\alpha ,\lambda
}]=\xi _{\alpha ,\lambda }$. 

The action of the elements $\Upsilon ( 0,b,0) \in \mathcal{A}( n)
$ of the\ \ homogeneous group on these representations is given
by the dual automorphisms 
\begin{equation}
\begin{array}{rl}
 \left. \left( {\widehat{\varsigma }}_{\Upsilon ( 0,b,0) }\xi _{\alpha
,\lambda }\right) \left( \Upsilon ( a,0,\iota ) \right) |\phi \right\rangle
& \left. =\xi _{\alpha ,\lambda }( \varsigma _{\Upsilon ( 0,b,0)
}\Upsilon ( a,0,\iota ) ) |\phi \right\rangle   \\
  & \left. =\xi _{\alpha ,\lambda }( \Upsilon ( a,0,\iota -a\cdot
b) ) |\phi \right\rangle   \\
  & =e^{i(  a\cdot \left( \alpha -\lambda  b\right) +\iota  \lambda
) }\overset{ }{\left. |\phi \right\rangle  } \\
  & =\xi _{\alpha -\lambda  b,\lambda }( \Upsilon ( a,0,\iota )
) \overset{ }{\left. |\phi \right\rangle  }.
\end{array}
\end{equation}

\noindent Therefore, the little group is the set of $\Upsilon (
0,b,0) \in \mathcal{K}\mbox{}^{\circ}$ that satisfy the fixed point
equation (14), 
\begin{equation}
{\widehat{\varsigma }}_{\Upsilon ( 0,b,0) }\xi _{ \alpha ,\lambda }=\xi
_{\alpha -\lambda  b,\lambda } =\xi _{\alpha ,\lambda }.
\end{equation}

\noindent The solution of the fixed point condition requires that
$\alpha -\lambda  b\equiv \alpha $. The $\lambda =0$ solution for
which the little group is $\mathcal{A}( n) $ is the degenerate case
corresponding to the homomorphism $\mathcal{H}( n) \rightarrow \mathcal{A}(
2n) $ with kernel $\mathcal{A}( 1) $ (See Appendix C, (123)).\ \ This
is just the abelian group that is not considered further here.\ \ \ The
faithful representation with $\lambda \neq 0$ requires $b=0$, and
therefore has the trivial little group $\mathcal{K}\mbox{}^{\circ}\simeq
\text{\boldmath $e$}\simeq \{\Upsilon ( 0,0,0) \}$.\ \ The stabilizer
is $\mathcal{G}\mbox{}^{\circ}\simeq \mathcal{A}( n+1) $.\ \ The
orbits are 
\begin{equation}
\mathbb{O}_{\lambda }=\left\{ {\widehat{\varsigma }}_{\Upsilon ( 0,b,0)
}[ \xi _{\alpha ,\lambda }] |b\in \mathbb{R}^{n}\right\} =\left\{
\xi _{b,\lambda }|b\in \mathbb{R}^{n}\right\} ,\ \ \lambda \in \mathbb{R}\backslash
\left\{ 0\right\} .
\end{equation}

All representations in the orbit are equivalent for the determination
of the semidirect product unitary irreducible representations. A
convenient representative of the equivalence class is $\xi _{0,\lambda
}$.\ \ The unitary representations $\sigma $ of the trivial little
group are trivial and therefore the representations of the stabilizer
are just $\varrho \mbox{}^{\circ}=\xi _{0,\lambda }$.\ \ The Hilbert
space ${\text{\boldmath $\mathrm{H}$}}^{\sigma }$ is\ \ also trivial
and therefore the Hilbert space of the stabilizer is ${\text{\boldmath
$\mathrm{H}$}}^{\varrho \mbox{}^{\circ}}={\text{\boldmath $\mathrm{H}$}}^{\sigma
}\otimes {\text{\boldmath $\mathrm{H}$}}^{\xi }\simeq \mathbb{C}.$

\subsubsection{Mackey induction}

\noindent The final step is to apply the Mackey induction theorem
to determine the faithful unitary irreducible representations of
the full $\mathcal{H}( n) $ group. The induction requires the definition
of the symmetric space
\begin{equation}
\mathbb{K}=\mathcal{G}/\mathcal{G}\mbox{}^{\circ}=\mathcal{H}( n)
/\mathcal{A}( n+1) \simeq \mathcal{A}( n) \simeq \mathbb{R}^{n},
\end{equation}

\noindent with the natural projection $\pi $ and a section $\Theta
$\ \ \ 
\begin{equation}
\begin{array}{l}
 \pi :\mathcal{H}( n)  \rightarrow \mathbb{K}:\Upsilon ( a,b,\iota
) \mapsto {\mathrm{k}}_{b}, \\
 \Theta :\mathbb{K}\rightarrow \mathcal{H}( n) :{\mathrm{k}}_{b}\mapsto
\Theta ( {\mathrm{k}}_{b}) =\Upsilon ( 0,b,0) .
\end{array}
\end{equation}

\noindent These satisfy $\pi ( \Theta ( {\mathrm{a}}_{b}) ) ={\mathrm{a}}_{b}$
and so $\pi \circ \Theta ={\mathrm{Id}}_{\mathbb{K}}$ as required.
Using (2), an element of the Weyl-Heisenberg group\ \ $\mathcal{H}(
n) $ can be written as, 
\begin{equation}
\Upsilon ( a,b,\iota ) =\Upsilon ( 0,b,0) \Upsilon ( a,0,\iota +\frac{1}{2}a\cdot
b) .
\end{equation}

\noindent The cosets are therefore defined by 
\begin{equation}
\begin{array}{rl}
 {\mathrm{k}}_{b} & =\left\{ \Upsilon ( 0,b,0) \Upsilon ( a,0,\iota
+\frac{1}{2}a\cdot b) |a\in \mathbb{R}^{n},\iota \in \mathbb{R}\right\}
\\
  & =\left\{ \Upsilon ( 0, b, 0) \mathcal{A}( n + 1) \right\} 
\end{array}
\end{equation}

\noindent Note that
\begin{equation}
\ \ \Upsilon ( a,b,\iota ) {\mathrm{k}}_{x} ={\mathrm{k}}_{x+b},\ \ \ \ x\in
\mathbb{R}^{n}.%
\label{PH: action of group on coset}
\end{equation}

\noindent The Mackey induced representation theorem can now be applied
straightforwardly.\ \ \ First, the Hilbert space is 
\begin{equation}
{\text{\boldmath $\mathrm{H}$}}^{\varrho }={\text{\boldmath $L$}}^{2}(
\mathbb{K},H^{\varrho \mbox{}^{\circ}}) \simeq {\text{\boldmath
$L$}}^{2}( \mathbb{R}^{n},\mathbb{C}) .
\end{equation}

\noindent Next the Mackey induction Theorem 7 yields
\begin{equation}
\psi ^{\prime }( {\mathrm{k}}_{x}) =\left( \varrho ( \Upsilon (
a,b,\iota ) )  \psi \right) \left( {\Upsilon ( a,b,\iota ) }^{-1}{\mathrm{k}}_{x}\right)
=\varrho \mbox{}^{\circ}( \Upsilon ( a \mbox{}^{\circ},0,\iota \mbox{}^{\circ})
) \psi ( {\mathrm{k}}_{x-b}) 
\end{equation}

\noindent Using the Weyl-Heisenberg group product (2),
\begin{equation}
\begin{array}{rl}
 \Upsilon ( \mathit{a\mbox{}^{\circ}},\mathit{b\mbox{}^{\circ}},\iota
\mbox{}^{\circ})  & ={\Theta ( {\mathrm{k}}_{x}) }^{-1}\Upsilon
( a,b,\iota )  \Theta ( {\Upsilon ( a,b,\iota ) }^{-1}{\mathrm{k}}_{x})
\\
  & =\Upsilon ( 0, -x, 0) \Upsilon ( a, b, \iota ) \Upsilon ( 0,
x - b, 0)  \\
  & =\Upsilon ( a,0,\iota +a\cdot \left( x-\frac{1}{2}b\right) )
.
\end{array}
\end{equation}

\noindent  We lighten notation using the isomorphism ${\mathrm{k}}_{x}\mapsto
x$.\ \ The induced representation theorem then yields
\begin{equation}
\begin{array}{rl}
 \psi ^{\prime }( x)  & =\xi _{0,\lambda }(\Upsilon ( a,0,\iota
+x\cdot a-\frac{1}{2}a\cdot b) \psi ( x-b)  \\
  & =e^{i \lambda ( \iota +x\cdot a-\frac{1}{2}a\cdot b) }\psi (
x-b) .
\end{array}%
\label{PH: WH UIR}
\end{equation}

\noindent Using Taylor expansion, we can write
\begin{equation}
\psi ( x-b) = e^{-b^{i} \frac{\partial }{\partial  {x}^{i}}}\psi
( x) .
\end{equation}

\noindent The\ \ Baker Campbell-Hausdorff formula \cite{Hall} enables
us to combine the exponentials
\begin{equation}
\psi ^{\prime }( x) =e^{i \left(  \lambda  \iota +\lambda  a^{i}
x_{i}+ b^{i}i\frac{\partial }{\partial x^{i}}\right) }\psi ( x)
=e^{i \left( a^{i}{\widehat{A}}_{i} +b^{i}{\widehat{B}}_{i} +\iota \widehat{I}\right)
}\psi ( x) .%
\label{PH: WH general nondegenerate representations}
\end{equation}

The representation of the algebra is therefore
\begin{equation}
\widehat{I}\psi ( x) =\lambda  \psi ( x)  ,\ \ \ {\widehat{A}}_{i}\psi (
x) =\lambda  x_{i}\psi ( x) ,\ \ \ {\widehat{B}}_{i}\psi ( x) =i \frac{\partial
}{\partial  x^{i}}\psi ( x) ,%
\label{PH: WH general algebra}
\end{equation}

\noindent that satisfies the commutation relations, $[{\widehat{B}}_{i},{\widehat{A}}_{j}]=i
\delta _{i,j}\widehat{I}$.\ \ This analysis can also be carried out
choosing $\Upsilon ( 0,b,\iota ) \in \mathcal{A}( n+1) $ to be the
elements of the normal subgroup and this yields the representation
with ${\widehat{B}}_{i}$ diagonal.\ \ 

\subsection{Unitary irreducible representations of the Hamilton
group}\label{PH: section: unitary rep Hamilton group}

Again, from Theorem 2, the projective representations of the Hamilton
group are equivalent to unitary representations of its cover. The
Hamilton group does not admit an algebraic extension and therefore
the central extension of the Hamilton group is its cover, $\widecheck{\mathcal{H}a}(
n) \simeq \overline{\mathcal{H}a}( n) $. The cover of the Hamilton
group may be written as a semidirect product group in the different
forms given in (12). 

The Mackey theorems may be used to compute the unitary irreducible
representations using any of these semi-direct product forms. We
use here the form with the nonabelian Weyl-Heisenberg group $\mathcal{H}(
n) $ as the normal subgroup.\ \ The reader can verify that the Mackey
theorem applied to the form with the abelian normal subgroup $\mathcal{A}(
n+1) $ gives the same result.\ \ \ 

For simplicity of exposition, we give the computation for the representations
of $\mathcal{H}a( n) $. Appendix D shows how the representation
of its cover $\overline{\mathcal{H}a}( n) $ is computed from these
results.

\subsubsection{Unitary representations of the Weyl-Heisenberg normal
subgroup}

We can apply the results of the previous section where we determined
the unitary irreducible representations of the Weyl-Heisenberg group.
In this section, the elements are $\Upsilon ( v,f,r) \in \mathcal{H}(
n) $ and\ \ the general element of the algebra is given in (29).
Choosing $\Upsilon ( v,0,r) $ to be the elements of the normal subgroup
$\mathcal{A}( n+1) $ of $\mathcal{H}( n) $, the representations
from (60) are
\begin{equation}
\varphi ^{\prime }( x) =\left( \xi ( \Upsilon ( v,f,r) )  \varphi
\right) \left( x\right) = e^{i \kappa ( r-\frac{1}{2}f\cdot v) +i
\kappa  v\cdot x}\varphi ( x-f) ,%
\label{PH: WH UIR subgroup of Ha}
\end{equation}

\noindent where $f,v,x\in \mathbb{R}^{n}, r,\kappa \in \mathbb{R},
\kappa \neq 0$ with $\mathrm{\varphi }\mathrm{\in }{\text{\boldmath
$\mathrm{H}$}}^{\xi }\simeq {\text{\boldmath $L$}}^{2}( \mathbb{R}^{n},\mathbb{C})
$. The hermitian representation of the basis $\{F_{i},G_{i},R\}$
of the Weyl-Heisenberg algebra are 
\begin{equation}
\widehat{R}=\xi ^{\prime }( R) =\kappa ,\ \ \ {\widehat{G}}_{i}=\xi ^{\prime
}( G_{i}) =\kappa  x_{i},\ \ \ {\widehat{F}}_{i}=\xi ^{\prime }( F_{i})
=-i \frac{\partial }{\partial  x^{i}}.%
\label{PH: rep WH algebra subgroup of Ha}
\end{equation}

\noindent Similar expressions result with the choice of elements
$\Upsilon ( 0,f,r) $ of the normal subgroup for which ${\widehat{F}}_{i}$
is diagonal.

\subsubsection{The $\rho$ representation}

The next step is to determine the stabilizer $\mathcal{G}\mbox{}^{\circ}$
and the representation $\rho $. It acts on the Hilbert space ${\text{\boldmath
$\mathrm{H}$}}^{\xi }$ and therefore the hermitian representations
$\rho ^{\prime }$ of the algebra of the little group must be realized
in the enveloping algebra of the Weyl-Heisenberg group. The hermitian
$\rho ^{\prime }$ representation of the basis of the algebra of\ \ the
Weyl-Heisenberg group are given by (65)\ \ as $\rho |_{\mathcal{H}(
n) }=\xi $.\ \ Define the hermitian $\rho ^{\prime }$ representation
generators as 
\begin{equation}
{\widehat{J}}_{i,j}=\rho ^{\prime }( J_{i,j}) =\frac{1}{\kappa }\left(
{\widehat{F}}_{i}{\widehat{G}}_{j}-{\widehat{F}}_{j}{\widehat{G}}_{i}\right) .%
\label{PH: Hamilton generator realization}
\end{equation}

The commutation relations for the generators in the $\rho ^{\prime
}$ representation are directly computed using\ \ (65)\ \ to be the
commutation relations for the Hamilton algebra given by (26) with
the $i$ inserted for the hermitian representation as noted in (41).

This is a $\rho ^{\prime }$ representation of the entire algebra
of the Hamilton group and therefore the stabilizer is the Hamilton
group itself.\ \ Using (22), the group action is given by
\begin{equation}
\rho ( \mathrm{R}( \theta ) ) \varphi ( x) =e^{-\theta ^{k,j}( x_{k}\frac{\ \ \partial
\ \ }{\partial  x^{j}}- x_{j}\frac{\partial \ \ }{\partial  x^{k}}
) }\psi ( x) = \varphi ( {\mathrm{R}}^{-1}( \theta )  x) .
\end{equation}

\noindent Using the semidirect group property (24), $\mathrm{K}(
\mathrm{R},v,f,r) =\Upsilon ( v,f,r)  \mathrm{R},$ where $\Upsilon
( v, f, r)  \simeq  \mathrm{K}( 1, v, f, r) $ and $\mathrm{R} \simeq
\mathrm{K}( \mathrm{R}, 0, 0, 0) $ and putting it together with
(20), the full $\rho $ representation is
\begin{equation}
\varphi ^{\prime }( x) =\left( \xi ( \Upsilon ( v,f,r) )  \rho (
\mathrm{R}) \varphi \right) \left( x\right) =e^{i \kappa ( r-\frac{1}{2}v\cdot
f+ v\cdot  x) }\varphi ( {\mathrm{R}}^{-1}x-f) .
\end{equation}

\noindent As the stabilizer is the full Hamilton group (and the
little group is $\mathcal{K}=\overline{\mathcal{S}\mathcal{O}}(
n) $), the Mackey theorem (Theorem 7) applies without requiring
use of the induced representation Theorem 6.

\subsubsection{Nondegenerate unitary irreducible representations
}

The faithful unitary irreducible representations of the Hamilton
group from the application of the Mackey theorem (Theorem 7) is
\begin{equation}
\varrho ( \mathrm{K}( \mathrm{R},v,f,r) )  = \sigma ( \mathrm{R})
\otimes  \rho ( \mathrm{K}( \mathrm{R},v,f,r) ) =\sigma ( \mathrm{R})
\otimes  \xi ( \Upsilon ( v,f,r) )  \rho ( \mathrm{R}) .%
\label{PH: Hamilton rep rho curly sigma rho}
\end{equation}

\noindent As the stabilizer is the group itself, induction is not
required.\ \ The $\sigma ( R) $ are the ordinary unitary irreducible
representations of $\overline{\mathcal{S}\mathcal{O}}( n) $\ \ that
act on a finite dimensional Hilbert space ${\text{\boldmath $\mathrm{H}$}}^{\sigma
}\simeq \mathbb{V}^{N}$.\ \ Therefore, the full representation acts
on the Hilbert space
\begin{equation}
 {\text{\boldmath $\mathrm{H}$}}^{\varrho }={\text{\boldmath $\mathrm{H}$}}^{\sigma
}\otimes {\text{\boldmath $\mathrm{H}$}}^{\xi }=\mathbb{V}^{N}\otimes
{\text{\boldmath $L$}}^{2}( \mathbb{R}^{n},\mathbb{C}) .\ \ 
\end{equation}

\noindent The full nondegenerate Hamilton representations are the
direct product given in (27).\ \ 
\begin{equation}
\begin{array}{rl}
 \varphi ^{\prime }( x)  & =\left( \varrho ( K( R,v,f,r) )  \varphi
\right) \left( x\right)  \\
  & =\left( \sigma ( \mathrm{R})  \otimes \xi ( \Upsilon ( v,f,r)
)  \rho ( \mathrm{R}) \varphi \right) \left( x\right) .
\end{array}
\end{equation}

In particular, for $n=3$, $\overline{\mathcal{S}\mathcal{O}}( 3)
=\mathcal{S}\mathcal{U}( 2) $ with $N=2j+1$ and the $\sigma $ representation
is given in terms of the well known $D^{j}( \mathrm{R}( \theta )
) $ representation matrices. For notation reasons, we set $x=\widetilde{f}$
as it is clear that it has the meaning of force with $\kappa $ having
the dimensions of the reciprocal of power, 
\begin{equation}
\begin{array}{rl}
 {\varphi ^{\prime }}_{\widetilde{m}}( \widetilde{f})  & ={D^{j}( \mathrm{R})
}_{\widetilde{m}}^{m} e^{i \kappa ( r-\frac{1}{2}v\cdot f+ v\cdot  \widetilde{f})
}\varphi _{m}( {\mathrm{R}}^{-1}\widetilde{f}-f) 
\end{array}.%
\label{PH: UIR Hamilton}
\end{equation}

The above representations use the choice of the normal subgroup
of $\mathcal{A}( n+1) $ of the Weyl-Heisenberg subgroup $\mathcal{H}(
n) $ that is generated by $\{G_{i},R\}$. These generators are diagonal
in the resulting representations (65) and (72). We could equally
well have chosen the normal subgroup to be generated by $\{F_{i},R\}$
resulting in similar representations with these generators diagonal
\begin{equation}
\begin{array}{rl}
 {{\widetilde{\varphi }}^{\prime }}_{m^{\prime }}( \widetilde{v})  & ={D^{j}(
\mathrm{R}) }_{m^{\prime }}^{m} e^{i \kappa  \left( r+\frac{1}{2}v\cdot
f+f\cdot  \widetilde{v}\right) }{\widetilde{\varphi }}_{m}( {\mathrm{R}}^{-1}\widetilde{v}-
v) 
\end{array}.%
\label{PH: UIR Hamilton v diag}
\end{equation}

The degenerate cases corresponding to the homomorphisms in Appendix
C (123) may be similarly computed.

\subsubsection{Casimir invariants}

The Casimir invariants for the case $n=3$ are given in (20).\ \ A
direct computation shows that $\rho ^{\prime }( C_{2}) =0$.\ \ Therefore,
\begin{equation}
\varrho ^{\prime }( C_{1}) =\rho ^{\prime }( C_{1}) =\rho ^{\prime
}( R) =\kappa ,\ \ \varrho ^{\prime }( C_{2}) ={\rho ^{\prime }(
R) }^{2}\sigma ^{\prime }( J^{2}) =\kappa ^{2}j( j+1) ,
\end{equation}

\noindent where $\kappa \in \mathbb{R}\backslash \{0\}$, and $j$
is half integral.\ \ The $(\kappa ,j)$ label the faithful irreducible
representations.\ \ Similar considerations apply for the degenerate
cases. 

\subsection{Unitary irreducible representations of the Galilei group}\label{PH:
section: Galilei group}

The projective representations of the inhomogeneous Euclidean group
are equivalent to the unitary representations of the cover of the
Galilei group.\ \ The Galilei group may be written as a semidirect
product in several different forms (31).\ \ Any of these forms may
be used in the Mackey theorems to determine the unitary irreducible
representations.\ \ The Mackey theorems for the form with the abelian
normal subgroup $\mathcal{A}( n+1) $ has been studied in reference
\cite{Voisin}.\ \ We choose here to use the form where $\mathcal{N}\simeq
\mathcal{H}( n) $ is the normal subgroup and the homogeneous group
is $\mathcal{K}\simeq \mathcal{A}( 1) \otimes \overline{\mathcal{S}\mathcal{O}}(
n) $.\ \ \ A general element of the Galilei group is
\begin{equation}
\Gamma ( \mathrm{R},t,v,q,s) \simeq \Upsilon ( v,q,s) \mathrm{A}(
t) \mathrm{R}
\end{equation}

\noindent where $\Upsilon ( v, q, s) \in \mathcal{H}( n) $, $\mathrm{A}(
t) \in \mathcal{A}( 1) $ and $\mathrm{R}\in \overline{\mathcal{S}\mathcal{O}}(
n) $.\ \ The general element of the algebra is given in (32). Note
that while the $\Upsilon $ are again elements of the Weyl-Heisenberg
group, in this case they are parameterized by position, velocity
and the the parameter of the central generator that is mass. 

 Again, the faithful unitary irreducible representation of elements
$\Upsilon ( v,q,s) \in \mathcal{H}( n) $ are 
\begin{equation}
\varphi ^{\prime }( \widetilde{p}) =\left( \xi ( \Upsilon ( v,q,s) )
\varphi \right) \left( u\right) = e^{\frac{i }{\hbar } \left( \mu
(  s-\frac{1}{2}q\cdot v ) +q\cdot \widetilde{p}\right) }\varphi ( \widetilde{p}-\mu
v) ,
\end{equation}

\noindent where $ q,\widetilde{p},v\in \mathbb{R}^{n}$, $s,\mu \in \mathbb{R}$,
$\mu \neq 0$\ \ and $\mathrm{\varphi }\mathrm{\in }{\text{\boldmath
$\mathrm{H}$}}^{\xi }\simeq {\text{\boldmath $L$}}^{2}( \mathbb{R}^{n},\mathbb{C})
$.\ \ The isomorphism (46) enables us to re-parameterize $s-\frac{1}{2}q\cdot
v\mapsto s$. The $\mu $ are the eigenvalues of the hermitian representation
of the central element $M$ of the basis $\{G_{i},P_{i},M\}$ of the
Weyl-Heisenberg algebra. The $\xi ^{\prime }$representation of these
generators, with momentum diagonal, are 
\begin{equation}
\widehat{M}=\xi ^{\prime }( M) =\mu ,\ \ {\widehat{P}}_{i}=\xi ^{\prime
}( P_{i}) = {\widetilde{p}}_{i},\ \ {\widehat{G}}_{i}=\xi ^{\prime }( G_{i})
=i \mu  \frac{\partial }{\partial {\widetilde{p}}^{i}}.%
\label{PH: rep WH algebra subgroup of Ga}
\end{equation}

\subsubsection{The $\rho$ representation}

 The next step is to determine the stabilizer $\mathcal{G}\mbox{}^{\circ}$
and the representation $\rho $. It acts on the Hilbert space ${\text{\boldmath
$\mathrm{H}$}}^{\xi }$ and therefore the hermitian representations
$\rho ^{\prime }$ of the little group must by realized in the enveloping
algebra of the Weyl-Heisenberg group \cite{Low}. The $\rho ^{\prime
}$ representation restricted to the normal subgroup are the $\xi
^{\prime }$ representation given in (77), $\rho |_{\mathcal{H}(
n) }=\xi $. The\ \ generators of the homogeneous group in the $\rho
^{\prime }$ representation with momentum diagonal are 
\begin{equation}
{\widehat{J}}_{i,j}=\rho ^{\prime }( J_{i,j}) =\frac{1}{\mu }\left(
{\widehat{P}}_{i}{\widehat{G}}_{j}-{\widehat{P}}_{j}{\widehat{G}}_{i}\right) ,\ \ \widehat{E}=\rho
^{\prime }( E) =\frac{1}{2 \mu }{\widehat{P}}^{2}.%
\label{PH: Galilei generator realization}
\end{equation}

\noindent These satisfy the commutation relations
\begin{equation}
\begin{array}{l}
 \left[ {\widehat{J}}_{i,j},{\widehat{J}}_{k,l}\right] =i( {\widehat{J}}_{j,k}
\delta _{i,l}+{\widehat{J}}_{i,l} \delta _{j,k}-{\widehat{J}}_{i,k} \delta
_{j,l}-{\widehat{J}}_{j,l} \delta _{i,k}) , \\
 \left[ {\widehat{J}}_{i,j},{\widehat{G}}_{k}\right] =i( {\widehat{G}}_{j} \delta
_{i,k}-{\widehat{G}}_{i}\delta _{j,k}) ,\ \ \left[ {\widehat{J}}_{i,j},{\widehat{P}}_{k}\right]
=i( {\widehat{P}}_{i} \delta _{j,k}-{\widehat{P}}_{j}\delta _{i,k}) , \\
 \left[ {\widehat{G}}_{i},\widehat{E}\right] =i{\widehat{P}}_{i},\ \ \ \left[
{\widehat{G}}_{i},{\widehat{P}}_{k}\right] =i\widehat{M} \delta _{i,k}.
\end{array}%
\label{PH: Ga algebra unitary rep}
\end{equation}

These are the commutation relations for the Galilei subgroup given
by (26-28) for the generators $\{J_{i,j},G_{i},P_{i},E,M\}$ with
the $i$ inserted for the hermitian representation as noted in (41).

\subsubsection{Nondegenerate unitary irreducible representations
}

The properties of the semidirect product enable us to write the
$\rho $ representation as
\begin{equation}
\rho (\Gamma ( \mathrm{R},t,v,q,s) = \xi ( \Upsilon ( v,q,s) ) \rho
( \mathrm{A}( t) ) \rho ( \mathrm{R}) .
\end{equation}

\noindent The $\sigma $ representations of $\mathrm{A}( t) \in \mathcal{A}(
1) $ are simply\ \ $\sigma ( \mathrm{A}( t) ) =e^{\frac{i}{\hbar
}t \sigma ^{\prime }( E) }=e^{\frac{i}{\hbar }t \varepsilon }$ with
$t,\varepsilon  \in \mathbb{R}$. We can then put everything together,
as in the Hamilton group case, to obtain the faithful unitary irreducible
representations 
\begin{equation}
\begin{array}{rl}
 \varphi ^{\prime }( \widetilde{p})  & =\left( \varrho ( \Gamma ( R,t,v,q,s)
)  \varphi \right) \left( \widetilde{p}\right)  \\
  & =\left( \sigma ( \mathrm{R}) \sigma ( \mathrm{A}( t) ) \otimes
\xi ( \Upsilon ( v,q,s) ) \rho ( \mathrm{A}( t) ) \rho ( \mathrm{R})
\varphi \right) \left( \widetilde{p}\right)  \\
  & =\sigma ( \mathrm{R}) e^{\frac{i}{\hbar }t \varepsilon }\otimes
e^{\frac{i}{\hbar }\left( \mu  s+ q\cdot p+ \frac{1}{2\mu }t p^{2}\right)
}\varphi ( {\mathrm{R}}^{-1}\widetilde{p}-\mu  v) .
\end{array}
\end{equation}

\noindent In this expression, $v,q,p\in \mathbb{R}^{n}$ and $s,\mu
\in \mathbb{R}$, $\mu \neq 0$ and $\varphi \in \mathbb{V}^{N}\otimes
{\text{\boldmath $L$}}^{2}( \mathbb{R}^{n},\mathbb{C}) .\ \ $\ \ Again,
for $n=3$, $N=2j+1$ and the $\sigma $ representation is given in
terms of the usual $D$ matrices,\ \ 
\begin{equation}
{\varphi ^{\prime }}_{\widetilde{m}}( \widetilde{p}) ={D^{j}( \mathrm{R})
}_{\widetilde{m}}^{m} e^{\frac{i}{\hbar }\left( \mu  s+ q\cdot p+t \left(
\varepsilon +\frac{1}{2\mu }p^{2}\right) \right) }\varphi _{m}(
{\mathrm{R}}^{-1}\widetilde{p}-\mu  v) .%
\label{PH: UIR Ga p}
\end{equation}

\noindent \ \ \ This is the same as the well known results for the
Galilei group determined from the abelian Mackey theorem (Theorem
8) using the semidirect product form in (31) with $\mathcal{A}(
n+2) $ as the normal subgroup \cite{Voisin}.

Appendix D shows how the representation of its cover $\overline{\mathcal{G}a}(
n) $ is computed from these results. The degenerate cases in Appendix
C\ \ (123) may similarly be computed using these methods.

\subsubsection{Casimir invariants}

The Casimir invariants of the Galilei group for $n=3$ are given
in (21). A straightforward calculation shows that $\rho ^{\prime
}( C_{2}) =0$ and $\rho ^{\prime }( C_{3}) =0$.\ \ Therefore,\ \ \ 
\begin{equation}
\begin{array}{l}
 \varrho ^{\prime }( C_{1}) =\rho ^{\prime }( M) =\mu , \\
 \varrho ^{\prime }( C_{2}) =2 \mu  \sigma ^{\prime }(  E) =2 \mu
\varepsilon , \\
 \varrho ^{\prime }( C_{3}) ={\mu  }^{2}\sigma ^{\prime }( J^{2})
={\mu  }^{2}j( j+1) .
\end{array}
\end{equation}

\noindent where $\mu ,\varepsilon \in \mathbb{R}$, $\mu \neq 0$
and $j$ half integral. The faithful\ \ irreducible representations
of the Galilei group are labeled by the eigenvalues $(\mu ,\varepsilon
,j)$. Similar considerations apply to the degenerate cases.

\subsection{Projective representations of the inhomogeneous Hamilton
group}\label{PH: section: Proj rep CE iha}

The projective representations of the inhomogeneous Hamilton group
$\mathcal{I}\mathcal{H}a( n) $ are the unitary representations of
its central extension\ \ $\mathcal{I}\widecheck{\mathcal{H}a}( n) \simeq
\overline{\mathcal{Q}\mathcal{H}a}( n) $ that was given in Section
3. We undertake the calculation for $\mathcal{Q}\mathcal{H}a( n)
$ and then show in Appendix D how the result for the cover follows.\ \ The
unitary irreducible representations obtained using the Mackey Theorem
7 consist of the nondegenerate faithful representations and the
rich set of degenerate representations\ \ that correspond to faithful
representations of the homomorphisms of the group given in Appendix
C (123). We focus here on the faithful representations and leave
the degenerate cases as an exercise for the reader using the same
methods.\ \ 

\subsubsection{The\ \ unitary irreducible representations of the
normal subgroup}

The normal subgroup of $\mathcal{Q}\mathcal{H}a( n) $ in the semidirect
product (19) is $\mathcal{N}\simeq \mathcal{H}( n+1) \otimes \mathcal{A}(
2) $.\ \ Its elements are the product of\ \ $\Upsilon ( q,p,t,\varepsilon
,\iota ) \in \mathcal{H}( n+1) $, where $q,p\in \mathbb{R}^{n}$,
$t,\varepsilon ,\iota \in \mathbb{R}$, and $\mathrm{A}( s,u) \in
\mathcal{A}( 2) $, with $s,u\in \mathbb{R}$. The faithful unitary
irreducible representations of the normal subgroup are just the
direct product of the unitary irreducible representations of $\mathcal{H}(
n+1) $ and $\mathcal{A}( 2) $
\begin{equation}
\begin{array}{rl}
 \psi ^{\prime } & =\widetilde{\xi }( \Upsilon ( q,p,t,\varepsilon ,\iota
) ) \widetilde{\xi }( \mathrm{A}( s,u) ) \psi  \\
  & =e^{i( \iota  \widetilde{I}+\frac{1}{\hbar }\left( q^{i}{\widetilde{P}}_{i}+p^{i}{\widetilde{Q}}_{i}+t\widetilde{E}-\varepsilon
\widetilde{T}\right) ) }e^{i( s\widetilde{M} +u\widetilde{A} ) }\psi .
\end{array}
\end{equation}

\noindent $\psi $ is an element of the Hilbert space ${\text{\boldmath
$\mathrm{H}$}}^{\xi }\simeq {\text{\boldmath $L$}}^{2}( \mathbb{R}^{n},\mathbb{C})
$ and we denote the hermitian representation of the generators with
tilde,\ \ $\widetilde{Z}={\widetilde{\xi }}^{\prime }( Z) $. The generators
$\{M,A,I\}$ are central and therefore their hermitian representation
is always diagonal.\ \ 
\begin{equation}
\widetilde{I}=\lambda ,\ \ \widetilde{M}=\mu ,\ \ \widetilde{A}=\alpha . 
\end{equation}

\noindent Any commuting subset of the hermitian representation of
the generators $\{{\widetilde{Q}}_{i},\widetilde{T},{\widetilde{P}}_{i},\widetilde{E}\}$\ \ may
be simultaneously diagonalized. Four canonical sets are $\{{\widetilde{P}}_{i},\widetilde{T}\}$,
$\{{\widetilde{Q}}_{i},\widetilde{T}\}$, $\{{\widetilde{P}}_{i},\widetilde{E}\}$,
and $\{{\widetilde{Q}}_{i},\widetilde{E}\}$.\ \ For example, if we diagonalize
$ \{{\widetilde{P}}_{i},\widetilde{T}\}$ the generators are realized in
the momentum-time representation,$ \psi ( \widetilde{p},\widetilde{t}) =\langle
\widetilde{p},\widetilde{t}|\psi \rangle $, as
\begin{equation}
{\widetilde{P}}_{i}={\widetilde{p}}_{i},\ \ \widetilde{T}=\lambda  \widetilde{t},\ \ {\widetilde{Q}}_{i}=-i
\lambda  \hbar \frac{\partial }{\partial {\widetilde{p}}_{i}},\ \ \widetilde{E}=i
\hbar \frac{\partial }{\partial \widetilde{t}}. %
\label{PH: WH Heisenberg commutation}
\end{equation}

These satisfy the Heisenberg commutation relations
\begin{equation}
\left[ {\widetilde{P}}_{i},{\widetilde{Q}}_{i}\right] =i \hbar  \widetilde{I}\delta
_{i,j},\ \ \ \ \left[ \widetilde{T},\widetilde{E}\right] =i \hbar  \widetilde{I}
\end{equation}

The Weyl-Heisenberg\ \ group representation in this basis is\ \ \ 
\begin{equation}
\begin{array}{rl}
 \psi ^{\prime }( \widetilde{p},\widetilde{t})  & \left. =\widetilde{\xi }(
\Upsilon ( q,p,t,\varepsilon ,\iota ) ) \widetilde{\xi }( \mathrm{A}(
s,u) ) \psi \right) \left( \widetilde{p},\widetilde{t}\right)  \\
  & =e^{i \vartheta }\begin{array}{l}
 \psi ( \widetilde{p}-p,\widetilde{t}-t) 
\end{array},
\end{array}%
\label{PH: WH rep Heisenberg group}
\end{equation}

\noindent where 
\begin{equation}
\vartheta =\left( s \mu +u \alpha +\lambda  \iota +\frac{1}{\hbar
} \left( q\cdot \widetilde{p}-\varepsilon  \widetilde{t}-\frac{\lambda }{2}\left(
q\cdot p-\varepsilon  t\right) \right) \right. 
\end{equation}

\noindent with $\mu ,\lambda \in \mathbb{R}\backslash \{0\}$ and
$ \widetilde{t}\in \mathbb{R}$, $\widetilde{p}\in \mathbb{R}^{n}$.\ \ The
physical Heisenberg commutation relations require $\lambda =1$ and
we therefore set $\lambda =1$ going forward. 

\subsubsection{The $\widetilde{\rho }$ representations}

The next step is to determine the representation $\widetilde{\rho }$
and the stabilizer $\mathcal{G}\mbox{}^{\circ}$ on which it acts
as defined in Theorem 7. (The tilde is to distinguish this $\rho
$ representation from the $\rho $ representation of the Hamilton
subgroup that we have already determined which will also be required
in this calculation.)\ \ The algebra of $\mathcal{H}a( n) $ may
be realized in the enveloping algebra of the algebra of $\mathcal{H}(
n+1) \otimes \mathcal{A}( 2) $.\ \ (In this section, the tilde on
a generator of the algebra denotes the ${\widetilde{\rho }}^{\prime
}$ representation, $\widetilde{Z}={\widetilde{\rho }}^{\prime }( Z) $.)
Note that ${\widetilde{\rho }}^{\prime }|_{\mathcal{N}}={\widetilde{\xi
}}^{\prime }$.
\begin{equation}
\begin{array}{ll}
 {\widetilde{J}}_{i,j}=\frac{1}{\lambda  \hbar }\left( {\widetilde{P}}_{j}{\widetilde{Q}}_{i}-{\widetilde{P}}_{i}{\widetilde{Q}}_{j}\right)
, & {\widetilde{R}}_{ }= \frac{1}{2 \lambda  \hbar }\left( {\widetilde{T}
}^{2}+\widetilde{M}\widetilde{A}\right) , \\
 {\widetilde{G}}_{i }=\frac{1}{\lambda  \hbar }\left(  \widetilde{T} {\widetilde{P}}_{i}-\widetilde{M}
{\widetilde{Q}}_{i}\right) , & {\widetilde{F}}_{i }=\frac{1}{\lambda  \hbar
}\left(  \widetilde{T} {\widetilde{Q}}_{i}+\widetilde{A} {\widetilde{P}}_{i}\right)
,
\end{array}
\end{equation}

The commutation relations for the generators in the ${\widetilde{\rho
}}^{\prime }$ representation may be directly computed and shown
to satisfy the algebra (26-28) of $\mathcal{Q}\mathcal{H}a( n) $
with the $i$ inserted for the hermitian representation as explained
in (41).

As all of the generators of $\mathcal{Q}\mathcal{H}a( n) $ are realized
in this ${\widetilde{\rho }}^{\prime }$ representation, the stabilizer
$\mathcal{G}\mbox{}^{\circ}$ is the entire group $\mathcal{Q}\mathcal{H}a(
n) $ and the {\itshape little} group is $\mathcal{H}a( n) $.\ \ Using\ \ the
properties of the semidirect product (22), the $\widetilde{\rho }$ representation
may be written as
\begin{equation}
\widetilde{\rho }( \Gamma ( \mathrm{R},v,f,r, q,t,p,\varepsilon ,\iota
,s,u) ) =\widetilde{\xi }( \Upsilon ( q,t,p,\varepsilon ,\iota ) ) \widetilde{\xi
}( \mathrm{A}( s,u) ) \widetilde{\rho }( \Upsilon ( v,f,r) ) \widetilde{\rho
}( \mathrm{R}) .
\end{equation}

In the momentum-time representation, the $\{{\widetilde{J}}_{i,j},{\widetilde{G}}_{i},{\widetilde{F}}_{i},\widetilde{R}\}$
generators are (with $\lambda =1$)
\begin{equation}
\begin{array}{ll}
 {\widetilde{J}}_{i,j}=i \left( {\widetilde{p}}_{i} \frac{\partial }{\partial
{\widetilde{p}}_{j}}-{\widetilde{p}}_{j}\frac{\partial }{\partial {\widetilde{p}}_{i}}\right)
, & {\widetilde{R}}_{ }= \frac{1}{\hbar }\left( {\widetilde{t} }^{2}+ \mu
\alpha \right) , \\
 {\widetilde{G}}_{i}=\frac{1}{\hbar }\left(  \widetilde{t} {\widetilde{p}}_{i}+
i \hbar  \mu  \frac{\partial }{\partial {\widetilde{p}}_{i}}\right)
, & {\widetilde{F}}_{i}=\frac{1}{\hbar }\left( \alpha  {\widetilde{p}}_{i}-i
\hbar  \widetilde{t}\frac{\partial }{\partial {\widetilde{p}}_{i}} \right)
.
\end{array}
\end{equation}

\noindent The $\widetilde{\rho }$ representation of the $\mathcal{S}\mathcal{O}(
n) $ subgroup with elements $\mathrm{R}$ is 
\begin{equation}
\psi ^{\prime }( \widetilde{p},\widetilde{t}) =\left( \widetilde{\rho }( \mathrm{R})
\psi \right) \left( \widetilde{p},\widetilde{t}\right) =e^{i \theta ^{i,j}{\widetilde{J}}_{i,j}}\psi
( \widetilde{p},\widetilde{t}) =\psi ( {\mathrm{R}}^{-1}\widetilde{p},\widetilde{t})
.
\end{equation}

For the\ \ $\widetilde{\rho }$ representation of the Weyl-Heisenberg
subgroup with elements $\Upsilon ( v, f, r) $, first note that a
general element of this algebra is 
\begin{equation}
\begin{array}{rl}
 \widetilde{Z} & =r\widetilde{R}+v^{i}{\widetilde{G}}_{i}+f^{i}{\widetilde{F}}_{i}
\\
  & =\frac{1}{\hbar }\left( r( {\widetilde{t} }^{2}+\mu  \alpha ) +\left(
v^{i}\widetilde{t} +f^{i}\alpha \right) {\widetilde{p}}_{i}+\left(  \mu
v^{i}-\widetilde{t} f^{i}\right)  \left( i \hbar  \frac{\partial }{\partial
{\widetilde{p}}_{i}}\right) \right) .
\end{array}
\end{equation}

Therefore,\ \ 
\begin{equation}
\begin{array}{rl}
 \psi ^{\prime }( \widetilde{p},\widetilde{t})  & =\left( \widetilde{\rho }(
\Upsilon ( v, f, r) ) \psi \right) \left( \widetilde{p},\widetilde{t}\right)
\\
  & =e^{i( v^{i}{\widetilde{G}}_{i}+f^{i}{\widetilde{F}}_{i}+r\widetilde{R})
}\psi ( \widetilde{p},\widetilde{t})  \\
  & =e^{i \vartheta ^{\prime }}\psi ( \widetilde{p}-\mu  v+t f,\widetilde{t})
\end{array}
\end{equation}

\noindent where 
\begin{equation}
\vartheta ^{\prime }=\frac{1}{\hbar }\left( r( {\widetilde{t} }^{2}+\alpha
\mu -\frac{1}{2}\left(  \mu  v-\widetilde{t} f\right) \cdot \left( v\widetilde{t}+\alpha
f\right) ) +\left( v \widetilde{t}+\alpha  f\right) \cdot \widetilde{p}\right)
.
\end{equation}

\noindent Putting together these equations with (88) gives the full
nondegenerate $\widetilde{\rho }$ representation of the\ \ group\ \ in
the momentum-time diagonal basis
\begin{equation}
\begin{array}{l}
 \begin{array}{l}
 \psi ^{\prime }( \widetilde{p},\widetilde{t}) =\left( \widetilde{\rho }( \Gamma
( \mathrm{R},v,f,r, q,t,p,\varepsilon ,\iota ,s,u) ) \psi \right)
\left( \widetilde{p},\widetilde{t}\right)  \\
 =e^{i( \vartheta +\vartheta ^{\prime }) }\psi ( {\mathrm{R}}^{-1}\widetilde{p}-\mu
v+t f-p,\widetilde{t}-t) .
\end{array}
\end{array}%
\label{PH: rho rep with p diagonal}
\end{equation}

 A similar calculation shows that in a position time basis, this
results in 
\begin{equation}
\begin{array}{l}
 \begin{array}{l}
 \psi ^{\prime }( \widetilde{q},\widetilde{t}) =\left( \widetilde{\rho }( \Gamma
( \mathrm{R},v,f,r, q,t,p,\varepsilon ,\iota ,s,u) ) \psi \right)
\left( \widetilde{q},\widetilde{t}\right)  \\
 =e^{\frac{i }{\hbar }\left( \widetilde{\vartheta }+{\widetilde{\vartheta
}}^{\prime }\right) }\psi ( {\mathrm{R}}^{-1}\widetilde{q}+\alpha  f+t
v-q,\widetilde{t}-t) 
\end{array}
\end{array}%
\label{PH: rho with q diag}
\end{equation}

\noindent where in this expression
\begin{equation}
\begin{array}{l}
 \widetilde{\vartheta }=s \mu +u \alpha + \iota +\frac{1}{\hbar } \left(
p\cdot \widetilde{q}-\varepsilon  \widetilde{t}+\frac{1}{2}\left( q\cdot
p-\varepsilon  t\right) \right) , \\
 {\widetilde{\vartheta }}^{\prime }=\frac{1}{\hbar }\left( r( {\widetilde{t}
}^{2}+\alpha  \mu +\frac{1}{2}\left(  \mu  v-\widetilde{t} f\right)
\cdot \left( v\widetilde{t} +\alpha  f\ \ \right) ) +\left( \mu  v-
\widetilde{t} f \right) \cdot \widetilde{q}\right) .
\end{array}
\end{equation}

\subsubsection{Nondegenerate unitary irreducible representations
}

As the stabilizer is the entire group $\mathcal{G}\mbox{}^{\circ}\simeq
\mathcal{Q}\mathcal{H}a( n) $, the Mackey induced representation
theorem (Theorem 6) is not required and the unitary irreducible
representations are given by 
\begin{equation}
\varrho ( \Gamma ) =\widetilde{\sigma }( \mathrm{K}) \otimes \widetilde{\rho
}( \Gamma ) =\sigma ( \mathrm{R}) \otimes \rho ( \mathrm{K}) \otimes
\widetilde{\rho }( \Gamma ) ,
\end{equation}

\noindent with $\mathrm{R}\in \mathcal{S}\mathcal{O}( n) $, $\mathrm{K}\in
\mathcal{K}\mbox{}^{\circ}\simeq \mathcal{H}a( n) $ and $\Gamma
\in \mathcal{Q}\mathcal{H}a( n) $.\ \ The $\widetilde{\sigma }$ representations
are the unitary irreducible representations of the Hamilton group
that are referred to as the $\varrho $ representations in (72),
$\widetilde{\sigma }( \mathrm{K}) \simeq \sigma ( \mathrm{R}) \otimes
\rho ( \mathrm{K}) $.\ \ The $\widetilde{\rho }$ representations are
given above in (97).\ \ \ 

Putting it all together, for $n=3$ the nondegenerate unitary irreducible
representation of $\mathcal{Q}\mathcal{H}a( 3) $ in a basis with
$\{G_{i},P_{i},T\}$ diagonal is
\begin{equation}
\begin{array}{l}
 \psi _{\widetilde{m},\widetilde{f}}^{\prime }( \widetilde{p},\widetilde{t}) = e^{i
\vartheta ^{\prime \prime \prime }}{D^{j}( \mathrm{R}) }_{\widetilde{m}}^{m}\begin{array}{l}
 \psi _{m,{\mathrm{R}}^{-1}\widetilde{f}-f }( {\mathrm{R}}^{-1}\widetilde{p}-\mu
v+f\widetilde{t}-p,\widetilde{t}-t) 
\end{array},
\end{array}%
\label{PH: UIR QHa psi p t}
\end{equation}

\noindent where we have set $\lambda =1$ and the phase\ \ $\vartheta
^{\prime \prime \prime }$\ \ is
\[
\vartheta ^{\prime \prime \prime }=\vartheta +\vartheta ^{\prime
}+\vartheta ^{{\prime\prime}},\ \ \vartheta ^{{\prime\prime}}=\kappa
( r-\frac{1}{2}v\cdot f+ v\cdot  \widetilde{f}) .
\]

\noindent $\vartheta ^{{\prime\prime}}$ is the phase of the $\rho
$ representation of the Hamilton subgroup that is given in (72).\ \ 

One could also choose to have $\{F_{i},P_{i},T\}$ to be diagonal.
Using (73), the position-time wave functions are
\begin{equation}
\begin{array}{l}
 {\widetilde{\psi }}_{\widetilde{m},\widetilde{v}}^{\prime }( \widetilde{q},\widetilde{t})
=e^{i {\widetilde{\vartheta }}^{\prime \prime \prime }}{D^{j}( \mathrm{R})
}_{\widetilde{m}}^{m}\begin{array}{l}
 {\widetilde{\psi }}_{m,{\mathrm{R}}^{-1}\widetilde{v}-v }( {\mathrm{R}}^{-1}\widetilde{q}-v\widetilde{t}
-\alpha  f-q,\widetilde{t}-t) .
\end{array}
\end{array}%
\label{PH: UIR QHa psi q t}
\end{equation}

\noindent where we have set $\lambda =1$ the phase ${\widetilde{\vartheta
}}^{\prime \prime \prime }$ is (73)
\[
{\widetilde{\vartheta }}^{\prime \prime \prime }=\vartheta +\vartheta
^{\prime }+{\widetilde{\vartheta }}^{{\prime\prime}},\ \ {\widetilde{\vartheta
}}^{{\prime\prime}}=\kappa ( r+\frac{1}{2}v\cdot f+ f\cdot  \widetilde{v})
.
\]

\noindent Other combinations of generators that can be simultaneously
diagonalized include $\{G_{i},Q_{i},T\}$ and $\{F_{i},Q_{i},T\}$.
Their representations follow from (98) together with (72) and (73).
The Hilbert space for all of these representations is
\begin{equation}
{\text{\boldmath $\mathrm{H}$}}^{\varrho }=\mathbb{V}^{2j+1}\otimes
{\text{\boldmath $L$}}^{2}( \mathbb{R}^{n},\mathbb{C})  \otimes
{\text{\boldmath $L$}}^{2}( \mathbb{R}^{n+1},\mathbb{C}) .
\end{equation}

The corresponding representation of the Lie algebra is
\begin{equation}
\begin{array}{l}
 \varrho ^{\prime }( J_{i,j}) = \sigma ^{\prime }( J_{i,j}) +{\widehat{J}}_{i,j}+{\widetilde{J}}_{i,j},
\\
 \varrho ^{\prime }( \left\{ G_{i},F_{i},R\right\} ) =\left\{ {\widehat{G}}_{i}+{\widetilde{G}}_{i},{\widehat{F}}_{i}+{\widetilde{F}}_{i},\widehat{R}+\widetilde{R}\right\}
, \\
 \varrho ^{\prime }( \left\{ Q_{i},P_{i},T,E,I,M,A\right\} ) = \left\{
{\widetilde{Q}}_{i},{\widetilde{P}}_{i},\widetilde{T},\widetilde{E},\widetilde{I},\widetilde{M},\widetilde{A}\right\}
.
\end{array}
\end{equation}

Appendix D shows how the representation of the cover $\overline{\mathcal{Q}\mathcal{H}a}(
n) $ is computed from these results.

\subsubsection{Casimir invariants}

The Casimir invariants are given in (16) for the case $n=3$.\ \ A
straightforward calculation shows that ${\widetilde{\rho }}^{\prime
}( C_{4}) =0$ and\ \ ${\widetilde{\rho }}^{\prime }( C_{5}) =0$.\ \ Combining
this with the corresponding results for the Hamilton group that
is the homogeneous group, the representations of these Casimirs
are
\begin{equation}
\begin{array}{l}
 \varrho ^{\prime }( C_{1}) ={\widetilde{\rho }}^{\prime }( I) =\lambda
,\ \ \ \varrho ^{\prime }( C_{2}) ={\widetilde{\rho }}^{\prime }( M)
=\mu ,\ \ \ {\widetilde{\rho }}^{\prime }( C_{3}) ={\widetilde{\rho }}^{\prime
}( A) =\alpha , \\
 \varrho ^{\prime }( C_{4}) ={\widetilde{\rho }}^{\prime }( I) \rho
^{\prime }( R) -{\widetilde{\rho }}^{\prime }( M A) =\kappa  \lambda
-\mu  \alpha , \\
 \varrho ^{\prime }( C_{5}) ={\left( \alpha  \mu -\kappa  \lambda
\right) }^{2}j( j+1) .
\end{array}
\end{equation}

\noindent Thus the $(\lambda ,\mu ,\alpha ,\kappa ,j)$ label the
nondegenerate projective representations of the quantum mechanical
Hamilton group for $n=3$.\ \ Again, the physical Heisenberg commutation
relations correspond to the irreducible representation with $\lambda
=1$

\section{Discussion}

This paper started with the observation that a most basic feature
of quantum mechanics is that physical states are rays that are equivalence
classes of states in a Hilbert space defined up to a phase.\ \ A
quote from Dirac later in his life exemplifies this.\ \ "So if one
asks what is the main feature of quantum mechanics, I feel inclined
now to say that it is not noncommutative algebra. It is the existence
of probability amplitudes which underlie all atomic processes. Now
a probability amplitude is related to experiment but only partially.
The square of the modulus is something that we can observe. That
is the probability which the experimental people get. But besides
that there is a phase, a number of modulus unity which we can modify
without affecting the square of the modulus. And this phase is all
important because it is the source of all interference phenomena
but its physical significance is obscure.'' \cite{Dirac 2} 

This physical requirement that the states in quantum mechanics are
rays requires projective representations of symmetry groups in quantum
mechanics rather than ordinary unitary representations. The projective
representations have the remarkable property that a set of theorem
that enables us to compute the representations for a general class
of connected Lie groups. The projective representations of the groups
discussed in the paper illustrate the power of these theorems and
these methods can be directly applied to a other connected Lie groups.\footnote{These
method's may also be applied to Lie groups that are not connected
but in this case the central extension may not be unique and therefore
they must be addressed on a case by case basis. }\ \ First, the
cornerstone Theorem 1\ \ states that a projective representation
of a connected group is equivalent to a projective representation
that is unitary and linear.\ \ Theorem 2 state further that a projective
representations are equivalent to the unitary representations of
its central extension. Central extensions are simply connected and
therefore Levi's Theorem 4 states that they are equivalent to a
semidirect product of a semi-simple subgroup and a solvable normal
subgroup.\ \ The unitary representations of the semi-simple groups
are generally known and the solvable group in the applications we
encounter turn out to be semidirect products of abelian groups.\ \ We
have shown how to compute the irreducible representations of the
solvable Weyl-Heisenberg group, and the $\mathcal{G}a( n) $, $\mathcal{H}a(
n) $ and $\mathcal{Q}\mathcal{H}a( n) \text{}$groups that have it
as a normal subgroup, using the Mackey Theorems 6-8. 

Furthermore, the form of the semidirect product is constrained by
the automorphism Theorem 5. We expect any physical symmetry to leave
invariant the Heisenberg commutation relations.\ \ That is, under
a symmetry transformation on phase space must result in 
\begin{equation}
i \hbar  \delta _{i,j }=\left[ {\widehat{P}}_{i}, {\widehat{Q}}_{j}\right]
=\varrho ( g) [ {\widehat{P}}_{i}, {\widehat{Q}}_{j}] {\varrho ( g) }^{-1}=\left[
{\widehat{P^{\prime }}}_{i}, {\widehat{Q^{\prime }}}_{j}\right]  ,
\end{equation}

\noindent with ${\widehat{P^{\prime }}}_{i}=\varrho ( g) {\widehat{P}}_{i}{\varrho
( g) }^{-1}$and ${\widehat{Q^{\prime }}}_{i}=\varrho ( g) {\widehat{Q}}_{i}{\varrho
( g) }^{-1}$ where $g$ is an element of the symmetry group $\mathcal{G}$.
Similar considerations hold for the energy-time commutation relations.\ \ This
requires $\mathcal{G}$ to be a subgroup of the automorphism group
of the Weyl-Heisenberg group.\ \ Therefore, the maximal symmetry
is $\mathcal{D}\mathcal{S}p( 2n+2) \simeq \mathcal{D}\otimes \mathcal{I}\mathcal{S}p(
2n+2) $ as the central extension of this group is the Weyl-Heisenberg
automorphism group. Furthermore, the projective representation of
this maximal symmetry results in the Heisenberg commutation relations
and the Hilbert space for these operators. 

 These mathematical theorems immediately give us the result that
the quantum mechanical phase leads directly to the noncommutative
algebra of quantum mechanics as stated by Dirac in the quote at
the beginning of this section. The symplectic homogenous group of
$\mathcal{D}\mathcal{S}p( 2n) $ constrains the central extension
of the abelian subgroup $\mathcal{A}( 2n+2) $ to be the Weyl-Heisenberg
group and therefore particular projective representations of this
abelian group are the unitary representations of the Weyl-Heisenberg
group.\ \ The physical meaning of the abelian group is translations
on extended phase space $\mathbb{P}\simeq \mathbb{R}^{2n+2}$ and
the hermitian representation of the Lie algebra corresponding to
the unitary representations of the Weyl-Heisenberg group are the
Heisenberg commutation relations.\ \ 

 Neither the Weyl-Heisenberg group nor the symplectic group have
any concept of an invariant time line element. This is a world before
any relativistic structure is present.\ \ There is no notion of
null cones, past and future, or causality. 

The relativistic structures of invariant time and mass line elements
may then be defined, 
\begin{equation}
d \tau ^{2}=d t^{2}-\frac{1}{c^{2}}d q^{2},\ \ d \mu ^{2}=\frac{1}{c^{2}}d
\varepsilon ^{2}-d p^{2}.
\end{equation}

These may be regarded as two degenerate line elements on the cotangent
of extended phase space, $T^{*}\mathbb{P}$.\ \ This now differentiates
key properties of time and energy from position and momentum. They
introduce the concepts of null cones and the notions of past and
future and causality.\ \ \ The invariance of these line elements
requires the subgroup $\mathcal{L}( 1,n) $\footnote{$\mathcal{L}(
1,n) $ is the connected subgroup of $\mathcal{O}( 1,n) $ that is
the full symmetry.} of $\mathcal{S}p( 2n+2) $.\ \ The invariance
of both of these line elements results in inertial states. This
allows the phase space to be {\itshape broken apart} into space-time
and energy-momentum space as this symmetry does not mix these degrees
of freedom. As we have noted in the introduction, special relativistic
quantum mechanics results from the projective representations of
$\mathcal{I}\mathcal{L}( 1,n) $.\ \ \ 

However, not all physical states are inertial. This leads us to
follow Born \cite{born1},\cite{born2} and combine the invariants
into a single nondegenerate invariant line element for time 
\begin{equation}
d {\widetilde{\tau }}^{2}=d t^{2}-\frac{1}{c^{2}}d q^{2}-\frac{1}{b^{2}}d
p^{2}+\frac{1}{b^{2}c^{2}}d \varepsilon ^{2}.
\end{equation}

Requiring invariance of this line element results in the subgroup
$\mathcal{U}( 1,n) $ of $\mathcal{S}p( 2n+2) $. $\mathcal{L}( 1,n)
$ is the inertial subgroup of $\mathcal{U}( 1, n) $. The constants
$\{c,b,\hbar \}$ define the dimensional basis where $b$, that has
dimensions of force, is a bound on the rate of change of momentum
just as $c$ is a bound on the rate of change of position. In this
basis, gravitational coupling is defined by the dimensionless constant
$\alpha _{G}$ where $G=\alpha _{G}c^{4}/b$.\ \ As $G$ and $c$ are
known, determining $\alpha _{G}$ or $b$ defines the other. These
effects are manifest only for forcesapproaching $b$, which if $\alpha
_{G}$ is anywhere near unity, is very large.\ \ 

A noninertial relativistic quantum theory results from the projective
representations of $\mathcal{I}\mathcal{U}( 1, n) $ with this definition
of invariant time that includes inertial and noninertial states
\cite{Low5,Low6,jarvis-1,Govaerts,jarvis2}.
These are given by the unitary representations of its central extension
$\overline{\mathcal{Q}}( 1,n) \simeq \overline{\mathcal{U}}( 1,
n) \otimes _{s}\mathcal{H}( n+1) $. The hermitian algebra of the
unitary representations of $\mathcal{H}( n+1) $ are the Heisenberg
commutation relations.\ \ 

The limits of the line elements for small forces relative to $b$
and small velocity relative to $c$ are\ \ \ 
\begin{equation}
d {\widetilde{\tau }}^{2}\operatorname*{\rightarrow }\limits_{b\rightarrow
\infty }d \tau ^{2}\operatorname*{\rightarrow }\limits_{c\rightarrow
\infty }d t^{2},
\end{equation}

\noindent with corresponding In\"on\"u-Wigner group contractions
\cite{inonu,Low9} 
\begin{equation}
\mathcal{I}\mathcal{U}( 1,n) \operatorname*{\rightarrow }\limits_{b\rightarrow
\infty }\mathcal{I}\mathcal{O}a( 1,n) \operatorname*{\rightarrow
}\limits_{c\rightarrow \infty }\mathcal{I}\mathcal{H}a( 1,n) .%
\label{PH: group contraction sequence}
\end{equation}

The group $\mathcal{I}\mathcal{O}a( 1,n) $ and its projective representations
is studied in \cite{Low9}. There exists a homomorphism of $\mathcal{I}\mathcal{O}a(
1,n) $ onto $\mathcal{I}\mathcal{L}( 1,n) $ and therefore the usual
representations of special relativistic quantum mechanics are a
degenerate representation. 

In this paper, we are focussed on the full $b,c\rightarrow \infty
$ limit group $\mathcal{I}\mathcal{H}a( 1,n) $ that is the symmetry
in the {\itshape nonrelativistic} regime \cite{Low7},\cite{Low8},\cite{rutwig}.\ \ The
projective representations are the unitary representations of its
central extension $\overline{\mathcal{Q}\mathcal{H}a}( n) $.

The first summary observation is that the projective representations
of $\mathcal{I}\mathcal{H}a( 1,n) $ contains precisely the unitary
representations of a Weyl-Heisenberg subgroup for which its algebra
is the physical Heisenberg commutation relations as is also the
case for both the $\mathcal{I}\mathcal{U}( 1,n) $ and $\mathcal{I}\mathcal{O}a(
1,n) $ groups. Simply by considering projective representations
of the symmetry $\mathcal{I}\mathcal{H}a( n) $ on phase space, we
obtain the noncommutative algebra of quantum mechanics. Again, just
as stated by Dirac in the quote at the beginning of this section,
the noncommutative structure arises because of the existence of
the quantum phase.\ \ All of these symmetry have an abelian subgroup
of translations on extended phase space that is parameterized by
position, time, momentum and energy degrees of freedom and yet result
in the expected wave functions that are functions of the eigenvalues
of commuting subsets.\ \ That is we, obtain wave functions of the
form $\psi ( p,t) $ (101) and $\psi ( q,t) $ (102),\ \ and not $\psi
( q,p,e,t) $ as, for example, is given by Wigner \cite{wigner 1932},\cite{Kim}
or Moyal \cite{Moyal} in their phase space formulations of quantum
mechanics. The Fourier transform between position and momentum representations
is just the unitary representation of the isomorphism on the Weyl-Heisenberg
group (46).

The central extension also contains the central generator $M$ that
is mass. This generator appears in precisely the form required for
the Galilei group to be the inertial subgroup. This central extension
embodies energy, $M c^{2}$. The Galilei group $\overline{\mathcal{G}a}(
n) $ is both a subgroup and a group homomorphic to $\overline{\mathcal{Q}\mathcal{H}a}(
n) $ as given in Appendix C (123).\ \ A consequence of Theorem 3
is that the complete set of unitary irreducible representations
of $\overline{\mathcal{Q}\mathcal{H}a}( n) $ includes the faithful
unitary irreducible representations of all of these homomorphic
groups as degenerate representations. Therefore, the unitary irreducible
representations of $\overline{\mathcal{Q}\mathcal{H}a}( n) $ include
the usual unitary irreducible representations of $\overline{\mathcal{G}a}(
n) $ corresponding to the inertial states. However,\ \ the quantum
mechanical Hamilton representations are on a larger Hilbert space
and include noninertial states not present in the representations
of the Galilei group.\ \ \ 

The group $\overline{\mathcal{Q}\mathcal{H}a}( n) $ has a third
central generator $A$ that has physical dimensions of the reciprocal
of tension. Studying the limits (110) carefully shows that it also
embodies energy, $A b^{2}$ just as mass embodies energy, $M c^{2}$.\ \ $A$
is as fundamental to the physical interpretation of this mathematical
theory as $M$ and $I$. The latter two are clearly very fundamental
and so if this symmetry has physical relevance, $A$ must also be
fundamental.\ \ This is a definitive test of the overall symmetry
- either physical phenomena corresponding to the presence of $A$
must be found to exist or an explanation of why its eigenvalue $\alpha
$ is zero must be provided. At this point, we can only note that,
as it embodies energy it will gravitate, and we know that we are
missing a lot of gravitational ``mass'' and ``energy'' in the current
standard theory that we refer to as dark.\ \ \ 

The expression (101) for the unitary irreducible representations
of $\mathcal{Q}\mathcal{H}a( n) $ may be written as
\begin{equation}
\begin{array}{l}
 \psi _{\widetilde{m},{\widetilde{f}}^{\prime }}^{\prime }( {\widetilde{p}}^{\prime
},{\widetilde{t}}^{\prime }) = e^{i \vartheta ^{\prime \prime \prime
}}{D^{j}( \mathrm{R}) }_{\widetilde{m}}^{m}\begin{array}{l}
 \psi _{m,\widetilde{f} }( \widetilde{p},\widetilde{t}) 
\end{array}
\end{array}
\end{equation}

\noindent where
\begin{equation}
{\widetilde{p}}^{\prime }=\mathrm{R}( {\widetilde{p}}^{\prime }+\mu  v-
f\widetilde{t}+p) , {\widetilde{t}}^{\prime }=\widetilde{t}+t, {\widetilde{f}}^{\prime
}=\mathrm{R}( \widetilde{f}+f) .
\end{equation}

These transformations of momentum and time are of the expected form
if force is constant. This is because the symmetry is global in
the treatment that we have provided. However, in general, we expect
these transformations to be local \cite{Low7}.\ \ That is,
\begin{equation}
d{\widetilde{p}}^{\prime }=\mathrm{R}( d{\widetilde{p}}^{\prime }+v( p,q,t)
d \mu  - f( p,q,t)  d\widetilde{t}+d p) .
\end{equation}

This will require the symmetry group to be gauged so that it is
local before attempting a full physical interpretation. This will
be the topic of a subsequent paper.

PDJ acknowledges the support of the Australian-American Fulbright
Foundation and staff and colleagues at the Department of Physics,
University of Texas at Austin as well as the Department of Statistics,\ \ University
of California, Berkeley, for visits as an Australian senior Fulbright
scholar, during which part of this work was undertaken. 

\appendix

\section{Projective representations and central extensions}

This appendix shows how Theorem 2 results in the understanding of
projective representations as a unitary representation up to a phase.\ \ \ Consider
a simply connected group $\overline{\mathcal{G}}$ with central extension
$\widecheck{\mathcal{G}}$ where the central extension includes a nontrivial
algebraic central extension. The algebra of $\widecheck{\mathcal{G}}$
is spanned by the\ \ generators $\{X_{a, }A_{\alpha }\}$ where the
$A_{\alpha }$ are central, $[X_{a},A_{\alpha }]=0$ and $[A_{\beta
},A_{\alpha }]=0$.\ \ As $\widecheck{\mathcal{G}}$ is always simply
connected, an $\text{}$element $\Gamma \in \widecheck{\mathcal{G}}$
may be expressed as
\begin{equation}
\Gamma ( x,a) =e^{a^{\alpha }A_{\alpha }}\gamma ( x) \ \ ,\ \ \ \gamma
( x)  =e^{ x^{a}X_{a}},\ \ 
\end{equation}
where $x\in \mathbb{R}^{r}, a\in \mathbb{R}^{m}$ with $n=r+m$.\ \ The
$\mathcal{A}( m) $ is the central subgroup of $\widecheck{\mathcal{G}}$
with an algebra spanned by $A_{a}$. Using the Baker-Campbell-Hausdorff
formula, the group product is, 
\begin{equation}
\begin{array}{rl}
 \Gamma ^{\prime }( x^{\prime },a^{\prime })  \Gamma ( x,a)  & =e^{\left(
{a^{\prime }}^{\alpha }+a^{\alpha }\right) A_{\alpha }}e^{{\beta
( x^{\prime },x) }^{a}X_{a}+{\alpha ( x^{\prime },x) }^{\alpha }A_{\alpha
}} \\
  & =e^{\left( {a^{\prime }}^{\alpha }+a^{\alpha }+{\alpha ( x^{\prime
},x) }^{\alpha }\right) A_{\alpha }}\gamma ( x^{{\prime\prime}})
,
\end{array}
\end{equation}
where $x^{{\prime\prime}}=\beta ( x^{\prime },x) $. Furthermore,
the terms that define $\beta $ are precisely the Baker-Campbell-Hausdorff
terms that result from the group product `` $\cdot $ `` of $\overline{\mathcal{G}}$,
$\gamma ( x^{{\prime\prime}}) =\gamma ( x^{\prime }) \cdot \gamma
( x) $. Therefore, the group product may be written as
\begin{equation}
\Gamma ^{\prime }( x^{\prime },a^{\prime }) \Gamma ( x,a) =e^{\left(
{a^{\prime }}^{\alpha }+a^{\alpha }+ {\alpha ( x^{\prime },x) }^{\alpha
}\right) A_{\alpha }}\gamma ( x^{\prime })  \cdot \gamma ( x) .%
\label{PH: proj rep as phases}
\end{equation}

\noindent In this sense, the elements $\gamma ( x) \text{}$ with
the multiplication $\cdot $ may be regarded to be elements of $\overline{\mathcal{G}}$.
Note that $\overline{\mathcal{G}}$ is not necessarily a subgroup
of $\widecheck{\mathcal{G}}$ as the $\alpha $ depend on $x$ and $x^{\prime
}$.\ \ The map $\alpha :\overline{\mathcal{G}}\times \overline{\mathcal{G}}\rightarrow
\mathcal{A}( m) $ may be shown to be an element of the second cohomology
group $H^{2}( \mathcal{G},\mathcal{A}( m) ) $.

Theorem 1 states that every projective representation of any connected
Lie group is equivalent to a projective representation that is unitary
and therefore, up to an equivalence, we need only consider unitary
representations. The unitary representations of elements $\Gamma
( x,a) $ of\ \ $\widecheck{\mathcal{G}}$ may be written as\ \ $\varrho
( \Gamma ) =\varrho ( \gamma ( x) ) \ \ e^{i a\cdot \nu }$ with
$\nu \in \mathbb{R}^{m}$. The unitary representation of the abelian
central subgroup $\mathcal{A}( m) $ is the phase $ e^{i a\cdot \nu
}$. The unitary representations of this group product is
\begin{equation}
\varrho ( \Gamma ^{\prime }) \varrho (  \Gamma ) =e^{i( a^{\prime
}+a+ \alpha ( x^{\prime },x) ) \cdot \nu }\varrho ( \gamma ^{\prime
}( x^{\prime })  ) \cdot \varrho ( \gamma ( x) )  .%
\label{PH: rep up to a phase}
\end{equation}

\noindent This is a projective representation defined as a unitary
representations of $\overline{\mathcal{G}}$ up to a continuous phase
defined by the continuous central group $\mathcal{A}( m) $.\ \ Conversely,
given an expression of the form (119) where the elements $ \alpha
( x^{\prime },x) $ are elements of the second cohomology group,
then this is the unitary representation of a central extension (See
section 2.7 of \cite{Weinberg1}.\ \ If $\mathcal{G}$ is simply connected
so that $\mathcal{G}\simeq \overline{\mathcal{G}}$, then this is
the maximal set of phases that can be constructed. 

If $\mathcal{G}$ is not simply connected, we must consider the topological
central extension resulting from a nontrivial homotopy. As $\overline{\mathcal{G}}$
is simply connected, its homotopy group is trivial and\ \ any loop
can be continuously deformed into the null loop. The kernel of the
projection $\pi \mbox{}^{\circ}:\overline{\mathcal{G}}\rightarrow
\mathcal{G}:\gamma \mapsto \widetilde{\gamma }$ is the homotopy group
$\ker  \pi \mbox{}^{\circ}\simeq \mathbb{A}\subset \overline{\mathcal{G}}$
that is a discrete central subgroup. The continuous curves
\begin{equation}
\ \ \gamma :\left[ 0,1\right] \rightarrow \overline{\mathcal{G}}
,\ \ \ \gamma _{a}( 0) =e ,\ \ \gamma _{a}( 1)  = \mathrm{a},\ \ 
\mathrm{a}\in \mathbb{A},
\end{equation}

\noindent project onto loops in $\mathcal{G}$, as $\pi \mbox{}^{\circ}(
a) =e$, that are representatives of the homotopy classes.\ \ If
the there is a point $t_{1}\in [0,1]$ such that $\gamma ( t_{1})
={\mathrm{a}}_{1}$, then the projected loop winds twice, and if
there are $n$-1 such points, $\gamma ( t_{i}) ={\mathrm{a}}_{i}$\ \ the
projected loop winds $n$ times. If we consider the case $\gamma
( t) =\gamma _{1}( t) \gamma _{2}( t) {\gamma _{3}( t) }^{-1}$ with
$\text{}\gamma _{i}( 0) =e, \gamma _{i}( 1) =\gamma _{i}$,\ \ then
this has the property that $\gamma _{1}\gamma _{2}{\gamma _{3}}^{-1}=\mathrm{a}$\ \ projects
onto a loop where ${\widetilde{\gamma }}_{1}{\widetilde{\gamma }}_{2}{{\widetilde{\gamma
}}_{3}}^{-1}=e$ . Thus the product ${\widetilde{\gamma }}_{1}{\widetilde{\gamma
}}_{2}={\widetilde{\gamma }}_{3}$ is associated with a homotopy class
that in the covering group is just
\begin{equation}
\gamma _{1}\gamma _{2}=\mathrm{a} \gamma _{3}.%
\label{PH: covering group product}
\end{equation}

The unitary representation of (119) is $\varrho ( \gamma _{1}) \varrho
( \gamma _{2}) =e^{i \phi }\varrho ( \gamma _{3}) $ where $e^{i
\phi } = \varrho ( \mathrm{a}) $ are the discrete phases that are
the unitary representations of the discrete center $ \mathbb{A}$.
This again has the form of a unitary representation up to a phase
and is the maximal set arising from a discrete central group. However,
these phases due to the homotopy cannot be realized locally in terms
of the algebra.\ \ This is a nonlocal effect and it does not appear
in the local expression for the {\itshape representation up to a
phase} given in (117).\ \ \ 

As the full central extension has a maximal center $\mathcal{Z}\simeq
\mathbb{A}\otimes \mathcal{A}( m) $, these are the maximal set of
phases that can be so constructed.

\section{Matrix realization of $\mathcal{Q}\mathcal{H}a( n) $}

The group $\mathcal{Q}\mathcal{H}a( n) $ is a matrix group with
elements $\Gamma ( \mathrm{R},v,f,r,t,q,p,\varepsilon ,s,u,\iota
) $ realized by the $2n+6$ dimensional matrices 
\begin{equation}
\left( \begin{array}{llllllll}
  \mathrm{R} & 0 & 0 & f & 0 & 0 & 0 & p \\
 0 &  \mathrm{R} & 0 & v & 0 & 0 & 0 & q \\
 v^{\mathrm{t}} \mathrm{R} & -f^{\mathrm{t}}\mathrm{R} & 1 & r &
0 & 0 & 0 & e \\
 0 & 0 & 0 & 1 & 0 & 0 & 0 & t \\
 0 & v^{\mathrm{t}}\mathrm{R} & 0 & v^{2}/2 & 1 & 0 & 0 & s \\
 f^{\mathrm{t}}\mathrm{R} & 0 & 0 & f^{2}/2 & 0 & 1 & 0 & u \\
 {\left( q- t v\right) }^{\mathrm{t}}\mathrm{R} & -{\left( p- t
f\right) }^{\mathrm{t}}\mathrm{R} & -t & \widetilde{\varepsilon } &
0 & 0 & 1 & 2 \iota  \\
 0 & 0 & 0 & 0 & 0 & 0 & 0 & 1
\end{array}\right) %
\label{PH: QHa matrix}
\end{equation}

\noindent where $\widetilde{\varepsilon }=\varepsilon  -r t+ q^{\mathrm{t}}\mathrm{R}
f- p^{\mathrm{t}}\mathrm{R} v$.

There is no direct algorithm to find matrix groups. However, once
the matrices realizing the group elements are found, it is straightforward
to establish that they indeed are a matrix group. The above realization
was inferred by first noting that $\mathcal{H}a( n) \otimes _{s}\mathcal{H}(
n+1) \subset \mathcal{H}\mathcal{S}p( 2n+2) $ where
\begin{equation}
\mathcal{H}\mathcal{S}p( 2n+2) \simeq \mathcal{S}p( 2n+2) \otimes
_{s}\mathcal{H}( n+1) .
\end{equation}

\noindent $\mathcal{H}\mathcal{S}p( 2m) $ is known to be a matrix
group with elements that are $2m+2$ dimensional matrices of the
form
\begin{equation}
\left( \begin{array}{lll}
 \ \ \ \ \ A & 0 & w \\
 \ \ \ \ \ 0 & 1 & \iota  \\
 -w^{\mathrm{t}}\cdot \zeta  \cdot A & 0 & 1
\end{array}\right) ,\ \ \ A\in \mathcal{S}p( 2m) ,\ \ w\in \mathbb{R}^{m},\ \ \iota
\in \mathbb{R},\text{}
\end{equation}

\noindent where $\zeta $ is a symplectic matrix.\ \ Furthermore,
the matrix realization of the Galilei group has elements that are
$n+3$ dimensional matrices with the well known form
\begin{equation}
\left( \begin{array}{llll}
  \mathrm{R} & v & 0 & q \\
 0 & 1 & 0 & t \\
 v^{\mathrm{t}}\mathrm{R} & v^{2}/2 & 1 & s \\
 0 & 0 & 0 & 1
\end{array}\right) ,\ \ \ \ \mathrm{R}\in \mathcal{S}\mathcal{O}(
n) ,\ \ v,q\in \mathbb{R}^{n},\ \ t,s\in \mathbb{R}.
\end{equation}

These two facts enable us to deduce the above realization of $\mathcal{Q}\mathcal{H}a(
n) $. It is straightforward to verify that matrix multiplication
realizes the group product and inverse (20-21).\ \ Furthermore,
the derivative at the identity yields a matrix realization of the
algebra that satisfies the commutation relations (26-28).

\section{Homomorphic groups and degenerate representations}

The following is a table of the groups homomorphic to the Weyl-Heisenberg,
Hamilton and Galilei groups.\ \ The generators of the homomorphic
groups are the complement set of generators. That is, the union
of the set of generators that are a basis of the normal subgroup\ \ and
the set that is a basis of the homomorphic group is a basis for
the full group. 
\begin{table}[h]
\begin{tabular}[c]{l|lll}
Group & Normal & Homomorphic & Generators of\\
 & Subgroup & Group & Normal Subgroup \\\hline $\mathcal{H}( n)$  & $\mathcal{A}( 1)$  &
$\mathcal{A}( 2n)$  &
$\left\{I\right\}$  \\
   &   &   &   \\
$\overline{\mathcal{H}a}( n)$  & $\mathcal{A}( 1)$  &
$\overline{\mathcal{S}\mathcal{O}}(
n) \otimes _{s}\mathcal{A}( 2n)$  & $\left\{ R\right\}$  \\
   & $\mathcal{A}( n+1)$  & $\overline{\mathcal{S}\mathcal{O}}( n)
\otimes _{s}\mathcal{A}( n)$  & $\left\{ G_{i},R\right\}$  \\
   & $\mathcal{A}( n+1)$  & $\overline{\mathcal{S}\mathcal{O}}( n)
\otimes _{s}\mathcal{A}( n)$  & $\left\{ F_{i},R\right\}$  \\
   & $\mathcal{H}( n)$  & $\overline{\mathcal{S}\mathcal{O}}( n)$  &
$\left\{ G_{i},F_{i},R\right\}$  \\
   &   &   &   \\
$\overline{\mathcal{G}a}( n)$  & $\mathcal{A}( 1)$  &
$\overline{\mathcal{E}}(
n) \otimes _{s}\mathcal{A}( n+1)$  & $\left\{ M\right\}$  \\
   & $\mathcal{A}( n+1)$  & $\left( \overline{\mathcal{S}\mathcal{O}}(
n) \otimes \mathcal{A}( 1) \right) \otimes _{s}\mathcal{A}( n)$
& $\left\{ P_{i},M\right\}$  \\
   & $\mathcal{A}( n+2)$  & $\overline{\mathcal{E}}( n)$  & $\left\{
E,P_{i},M\right\}$  \\
   & $\mathcal{H}( n)$  & $\overline{\mathcal{S}\mathcal{O}}( n) \otimes
\mathcal{A}( 1)$  & $\left\{ G_{i},P_{i},M\right\}$  \\
   & $\mathcal{A}( 1) \otimes _{s}\mathcal{H}( n)$  & $\overline{\mathcal{S}\mathcal{O}}(
n)$  & $\left\{ E,G_{i},P_{i},M\right\}$
\end{tabular}
\label{PH: Degenerate cases H Ha Ga}
\end{table}

In addition, there is a homomorphism $\overline{\mathcal{S}\mathcal{O}}(
n) \rightarrow \mathcal{S}\mathcal{O}( n) $ with a kernel that is
the normal subgroup $\mathbb{Z}_{2}$. Therefore,\ \ for each of
the entries above containing an $\overline{\mathcal{S}\mathcal{O}}(
n) $, there is a corresponding homomorphic group with kernel $\mathbb{Z}_{2}$
that has an isomorphic Lie algebra.\ \ This defines a corresponding
set of homomorphic groups that is the same as the above with the
bars denoting the cover\ \ removed.\ \ These groups are not simply
connected and their fundamental homotopy\ \ group is $\mathbb{Z}_{2}$.\ \ From
Theorem 3, these appear as degenerate representations.

The following is a table of the groups homomorphic to $\mathcal{I}\widecheck{\mathcal{H}}a(
n) \simeq \overline{\mathcal{Q}\mathcal{H}}( n) $\ \ with a connected
normal subgroup.
\begin{table}[h]
\begin{tabular}[c]{lll}
 Normal & Homomorphic & Generators of\\
 Subgroup & Group & Normal Subgroup \\\hline
$\mathcal{A}( 1)$  & $\overline{\mathcal{H}a}( n) \otimes
_{s}\mathcal{H}(
n+1)$  & $\left\{ A\right\} ,\left\{ M\right\}$  \\
$ \mathcal{A}( 1)$  & $\overline{\mathcal{H}a}( n) \otimes
_{s}\mathcal{A}(
2n+2)$  & $\left\{ I\right\}$  \\
$\mathcal{A}( 2)$  & $\overline{\mathcal{H}a}( n) \otimes
_{s}\mathcal{H}(
n+1)$  & $\left\{ A,M\right\}$  \\
$\mathcal{A}( 2)$  & $\overline{\mathcal{H}a}( n) \otimes
_{s}\mathcal{A}(
2n+1)$  & $\left\{ A,I\right\} ,\left\{ M,I\right\}$  \\
$\mathcal{A}( 3)$  & $\overline{\mathcal{H}a}( n) \otimes
_{s}\mathcal{A}(
2n)$  & $\left\{ I,A,M\right\}$  \\
$\mathcal{A}( n+3)$  & $\overline{\mathcal{H}a}( n) \otimes
_{s}\mathcal{A}(
n+1)$  & $\left\{ P_{i},T,M,I\right\}$  \\
$\mathcal{A}( n+3)$  & $\overline{\mathcal{H}a}( n) \otimes
_{s}\mathcal{A}(
n+1)$  & $\left\{ Q_{i},T,A,I\right\}$  \\
$\mathcal{H}( n) \otimes \mathcal{A}( 2)$  &
$\overline{\mathcal{G}a}(
n)$  & $\left\{ P_{i},G_{i},R,T,M,I\right\}$  \\
$\mathcal{H}( n) \otimes \mathcal{A}( 2)$  &
$\overline{\mathcal{G}a}(
n)$  & $\left\{ Q_{i},F_{i},R,T,A,I\right\}$  \\
$\mathcal{A}( n+4)$  & $\overline{\mathcal{H}a}( n) \otimes
_{s}\mathcal{A}(
n)$  & $\left\{ P_{i},T,M,A,I\right\}$  \\
$\mathcal{A}( n+4)$  & $\overline{\mathcal{H}a}( n) \otimes
_{s}\mathcal{A}(
n)$  & $\left\{ Q_{i},T,A,M,I\right\}$  \\
$\mathcal{H}( n+1)$  & $\overline{\mathcal{H}a}( n) \otimes
_{s}\mathcal{A}(
2)$  & $\left\{ I,P_{i},Q_{i},E,T\right\}$  \\
$\mathcal{H}( n+1) \otimes \mathcal{A}( 2)$  &
$\overline{\mathcal{H}a}(
n)$  & $\left\{ I,P_{i},Q_{i},E,T,A,M\right\}$  \\
$\mathcal{H}( n+1) \otimes \mathcal{A}( 3)$  &
$\overline{\mathcal{S}\mathcal{O}}( n) \otimes \mathcal{A}( 2n)$ &
$\left\{ I,P_{i},Q_{i},E,T,A,M,R\right\}$
\\
$\mathcal{H}( n+1) \otimes \mathcal{A}( n+3)$  &
$\overline{\mathcal{E}}(
n)$  & $\left\{ I,P_{i},Q_{i},E,T,A,M,F_{i,}R\right\}$  \\
$\mathcal{H}( n+1) \otimes \mathcal{A}( n+3)$  &
$\overline{\mathcal{E}}(
n)$  & $\left\{ I,P_{i},Q_{i},E,T,A,M,G_{i,}R\right\}$  \\
$\mathcal{H}( n+1) \otimes \mathcal{H}( n) \otimes \mathcal{A}(
2)$ & $\overline{\mathcal{S}\mathcal{O}}( n)$  & $\left\{
I,P_{i},Q_{i},E,T,A,M,F_{i},G_{i,}R\right\}$
\end{tabular}
\label{PH: Homomorphisms of CE IH}
\end{table}

Again, there are also the homomorphisms resulting from $\overline{\mathcal{S}\mathcal{O}}(
n) \rightarrow \mathcal{S}\mathcal{O}( n) $. Noting Theorem 3, this
shows that the\ \ projective representations of the inhomogeneous
Hamilton group has a rich set of degenerate representations.

\section{Representations of the cover of the groups for $n=3$}

The $\mathcal{H}a( n) $, $\mathcal{G}a( n) $ and $\mathcal{Q}\mathcal{H}a(
n) $ all have a $\mathcal{S}\mathcal{O}( n) $ subgroup appearing
in the semidirect product. The full central extension requires the
cover of these groups that have a corresponding $\overline{\mathcal{S}\mathcal{O}}(
n) $ subgroup.\ \ The manner of obtaining these cases follows the
method of determining the cover of the Euclidean group $\mathcal{E}(
n) $.

The terms of the form $\mathrm{R} x$,\ \ with $\mathrm{R}\in \mathcal{S}\mathcal{O}(
n) $ realized by $n$ dimensional real orthogonal matrices and $x\in
\mathbb{R}^{n}$ where $x$ is one of $v,f,p,q$ are actually automorphisms
of the form 
\begin{equation}
\varsigma _{\Gamma ( R,0,..0) }\Gamma ( 1,..,x,...) =\Gamma ( 1,...,\mathrm{R}
x,...) .
\end{equation}

For $n=3$, $\overline{\mathcal{S}\mathcal{O}}( 3) =\mathcal{S}\mathcal{U}(
2) $ with a 2:1 homomorphism $\pi :\mathcal{S}\mathcal{U}( 2) \rightarrow
\mathcal{S}\mathcal{O}( 3) : \overline{\mathrm{R}}\mapsto \mathrm{R}$.
Elements $\overline{\mathrm{R}}\in \mathcal{S}\mathcal{U}( 2) $
are realized by 2 dimensional complex unitary matrices.\ \ The automorphism
action on $x$ is given by representing the $x$ as the 2 dimensional
hermitian matrices $\overline{x}=x^{i}\sigma _{i}$ where the $\sigma
_{i}$ are the Pauli matrices.\ \ \ Then\ \ 
\begin{equation}
\varsigma _{\Gamma ( \overline{\mathrm{R}},0,..0) }\Gamma ( 1,..,\overline{x},...)
=\Gamma ( 1,...,\overline{\mathrm{R}} \overline{x}{\overline{\mathrm{R}}}^{-1},...)
.
\end{equation}

Substituting $\overline{\mathrm{R}} \overline{x}{\overline{\mathrm{R}}}^{-1}$
into the expressions where $\mathrm{R} x$ appears and allowing the
$j$ labeling the $D^{j}$ matrices to take half integral values gives
the representation for the cover of the groups. \label{con}\label{lcon}\label{coverl}\label{dq}\label{disc}\label{gqt}\label{spn}\label{tr}\label{ceh}\label{dot}\label{cas}\label{ncon}\label{cs}

\end{document}